\documentclass[12pt]{article}
\pdfoutput=1

\usepackage{amsmath,amssymb,jheppub,graphicx}
\usepackage{subfig}
\usepackage{centernot}
\usepackage{mciteplus}
\usepackage{color}
\allowdisplaybreaks[1]
\usepackage{tikz}
\usepackage{hyperref}
\usepackage{lipsum}
\usepackage{enumitem}
\tikzset{node distance=2cm, auto}

\usepackage{bm}

\newcommand{\be}{\begin{equation}}
\newcommand{\ee}{\end{equation}}
\newcommand{\bea}{\begin{eqnarray}}
\newcommand{\eea}{\end{eqnarray}}

\newcommand{\todo}[1]{}
\renewcommand{\todo}[1]{{\color{red} TODO: {#1}}}

\def\I{{\cal I}}
\def\R{{\mathbb R}}
\def\coeff#1#2{{\textstyle {\frac {#1}{#2}}}}

\def\half{\coeff 12}

\begin{document}
\baselineskip=15pt

\title{Critical Points at Infinity, Non-Gaussian Saddles, and Bions} 

\author[a,b]{Alireza Behtash,}
\author[a,c]{Gerald V. Dunne,}
\author[b]{Thomas Sch\"afer,}
\author[a,d]{Tin Sulejmanpasic,}
\author[a,b]{Mithat \"Unsal}

\affiliation[a]{Kavli Institute for Theoretical Physics, University of 
California, Santa Barbara, CA 93106, USA}
\affiliation[b]{Department of Physics, North Carolina State University, 
Raleigh, NC 27695, USA}
\affiliation[c]{Department of Physics, University of Connecticut, Storrs, 
CT 06269-3046, USA}
\affiliation[d]{Philippe Meyer Institute, Physics Department, \'Ecole 
Normale Sup\'erieure, 
PSL Research University, 24 rue Lhomond, F-75231 Paris Cedex 05, France}

\abstract{
It has been argued that many non-perturbative phenomena in quantum 
mechanics (QM) and quantum field theory (QFT) are determined by complex 
field configurations, and that these contributions should be understood 
in terms of Picard-Lefschetz theory.  In this work we compute the 
contribution from non-BPS multi-instanton configurations, such as 
instanton-anti-instanton $[{\mathcal I}\bar{\mathcal I}]$ pairs, and 
argue that these contributions should be interpreted as exact critical 
points at infinity. The Lefschetz thimbles associated with such critical 
points have a specific structure arising from the presence of non-Gaussian, 
quasi-zero mode (QZM), directions. When fermion degrees of 
freedom are present, as in supersymmetric theories, the effective bosonic 
potential can be written as the sum of a classical and a quantum potential. 
We show that in this case the semi-classical contribution of the critical 
point at infinity vanishes, but there is a non-trivial contribution that 
arises from its associated non-Gaussian QZM-thimble. This approach resolves 
several puzzles in the literature concerning the semi-classical contribution 
of correlated $[{\mathcal I}\bar{\mathcal I}]$ pairs. It has the surprising 
consequence that the configurations dominating the expansion of observables, 
and the critical points defining the Lefschetz thimble decomposition need 
not be the same, a feature not present in the traditional Picard-Lefschetz 
approach.

\medskip
\noindent
NSF-ITP-18-007}

\maketitle

\section{Introduction}
\label{sec_intro}

 The semi-classical expansion is an invaluable tool in quantum field theory 
and quantum mechanics \cite{Coleman:1978ae,ZinnJustin:2002ru}. However, 
beyond leading order, the naive semi-classical ``instanton gas'' expansion 
is typically ill-defined. The standard example of the problems that arise 
at higher order is that of an instanton-anti-instanton $[{\mathcal I}
\bar{\mathcal I}]$ pair. The instanton is a real saddle point of the path 
integral, characterized by a ``fugacity'' $e^{-S_I/g}$. But since instantons 
and anti-instantons attract one another, in the instanton gas framework 
there is no exact saddle point in the $[{\mathcal I}\bar{\mathcal I}]$ sector 
at finite separation. The action of such a configuration continuously 
decreases as the two pseudo-particles get closer. On the other hand, we 
expect physical observables, such as quantum mechanical energies, to have 
a well-defined expansion in powers of the instanton fugacity. Some progress 
was made by Bogomolny and Zinn-Justin a long time ago \cite{Bogomolny:1980ur,
ZinnJustin:1981dx,ZinnJustin:2002ru}, who computed higher order terms using 
analytic continuation, as reviewed below. However, this calculation left 
several conceptual and practical questions unanswered. In this paper, we 
resolve some of these issues, revisiting the BZJ analysis 
by taking advantage of recent progress in applying resurgence theory 
\cite{Cherman:2014ofa,Dunne:2012ae,Cherman:2013yfa,Misumi:2015dua,
Basar:2013eka,Argyres:2012ka,Anber:2014sda,Behtash:2015loa,Dunne:2016nmc,
Sulejmanpasic:2016fwr} and Picard-Lefschetz theory \cite{Kontsevich-3,
Kontsevich-1,Witten:2010zr,Witten:2010cx,Harlow:2011ny, Behtash:2015kva,
Behtash:2017rqj} to path integrals.
 
 It has become clear that the semi-classical expansion naturally lives in 
the \emph{complexified field space} \cite{Witten:2010cx, Witten:2010zr, 
Harlow:2011ny,Cherman:2014ofa,Behtash:2015kna,Behtash:2015kva, 
Behtash:2015zha,Behtash:2015loa,Kozcaz:2016wvy,
Fujimori:2016ljw,Dorigoni:2017smz,Nekrasov:2018pqq}.
The infinite dimensional space that appears in path integration is to be 
deformed into hypersurfaces in the complexified field space attached to
the saddles. Such hypersurfaces (known as ``Lefschetz thimbles'') should 
be chosen to guarantee that the integration always yields a convergent 
result. A convenient choice of these manifolds is generated by the 
{\it complexified gradient flow equations} in the field space  
\cite{Witten:2010zr,Witten:2010cx,fedoryuk} (as opposed to the real gradient 
flow equations or valley methods, cf. \cite{Schafer:1996wv}) given by
\be
\frac{\partial\phi}{\partial u}
  =\frac{\delta \bar S[\bar\phi]}{\delta\bar\phi} \, . 
\label{eq:flow}
\ee
Here, $\phi( x, u) $ is the (complexified) field that depends on the 
Euclidean space-time point $ x$, and the gradient \emph{flow time} $u$, 
and $S[\phi]$ is the holomorphic action functional.  The flow equation 
\eqref{eq:flow}  ensures that the real part of the action \emph{increases} 
along the flow, while the imaginary part remains constant, and hence naturally 
generalizes the stationary phase approximation in ordinary finite-dimensional 
integrals \cite{Pham,arnold}. If the flow equation is initialized in the 
vicinity of the saddle point\footnote{The flow equation clearly cannot 
be initialized exactly at the saddle point, since the RHS of \eqref{eq:flow} 
vanishes there.}, then the flow equations define a hypersurface with the 
two desired features: (1) the thimble contains the saddle point,  and (2) the 
path integral converges with the fields along the thimble.

 This observation implies that ``unstable'' saddles of the Euclidean 
action (i.e. critical points which are saddles rather than the minima of 
the action) should be treated on an \emph{equal footing} with those saddles 
that minimize the action. Historically, this point led to some confusion, 
because it was not always clear how to incorporate solutions with negative 
fluctuation modes. In principle, Picard-Lefschetz theory provides conceptual 
clarity in such situations. Note also that the contribution of ``unstable'' 
saddles generically requires the integration to be performed along complex 
manifolds in field space.

This issue arises unavoidably at  second (and higher) order in the 
semi-classical expansion. To describe physics at $O(e^{-2 S_I/g})$, the 
saddles $[{\mathcal I}{\mathcal I}]$, $[\bar{\mathcal I}\bar{\mathcal I}]$ 
and $[{\mathcal I}\bar{\mathcal I}]$ must all be treated on an equal 
footing. In the literature it is often asserted that instanton-anti-instanton 
configurations are not critical points of the action, because of the 
existence of a classical interaction between them of the form 
$V(\tau) \sim \pm \frac{1}{g} e^{-\tau}$, where $\tau$ is the quasi-zero 
mode direction. However, this potential has a critical point at 
infinity.\footnote{Interesting properties 
of critical points at infinity in ordinary integrals have been discussed 
in \cite{Pham}.}
In this work, by considering certain quantum mechanical (QM) systems on an 
Euclidean temporal circle with finite size $\beta$, we first find critical 
points at finite $\beta$. Upon taking $\beta\rightarrow\infty$ such saddles 
move to infinity. We show that the multi-instanton expansion, or cluster 
expansion, performed by using the Lefschetz thimbles associated with 
the QZM directions, resolves some important puzzles concerning the BZJ 
calculation of multi-instanton effects.  We argue  that the cluster 
expansion on the $\Gamma_{\rm QZM}$ thimble provides a conceptually complete 
framework for performing semi-classical expansions. 

 In our earlier studies \cite{Behtash:2015loa,Behtash:2015zha} we considered 
two prototypical QM examples: (i) systems with degenerate  harmonic minima 
coupled to $N_f$ Grassmann valued fields (corresponding to spin $(\half)^{N_f})$; 
and (ii) the related bosonic systems that arise after integrating out fermions,
which are characterized by non-degenerate harmonic minima. In fact, integrating 
out the Grassmann fields exactly, one obtains an effective quantum potential (as 
in ${\cal N}=1 $ supersymmetric quantum mechanics \cite{Witten:1982df,
Shifman:1995mm})
\begin{align} 
V(x) = v_0(x)  +   p\, g\,    v_1(x) \, , 
\label{eq:eff} 
\end{align}
where $v_0(x)$  is the classical potential, $p$ is a parameter related to 
$N_f$, and $v_1(x)$ is the potential generated by integrating out the fermions.
Note that the second term, involving $v_1(x)$,
 is a quantum correction, proportional to the coupling constant $g$. 
For $p\neq 0$, we refer to the potential $V(x)$ as the effective quantum 
potential. In  \cite{Behtash:2015loa,Behtash:2015zha} it was demonstrated that 
the bosonic theory governed by $V(x)$ can exhibit both real and complex bion 
configurations. Both of these configurations are {\it exact} solutions of 
the equations of motion of the {\it quantum action}, corresponding to the 
effective quantum potential in \eqref{eq:eff}.  In particular, the complex 
bion is not a solution of the ordinary (real)  classical equations of motion 
in the inverted potential (as described in standard textbooks, see 
\cite{Coleman:1978ae,ZinnJustin:2002ru}); rather it is an exact solution of the 
holomorphic classical equations of motion associated with the effective 
potential in \eqref{eq:eff}.  

In the works \cite{Balitsky:1985in,Behtash:2015kna,Behtash:2015kva, 
Behtash:2015zha,Behtash:2015loa,Buividovich:2015oju,Kozcaz:2016wvy,Fujimori:2016ljw} many 
arguments were made that indicate such complex solutions of the quantum 
action dominate the semi-classical analysis. But this observation, while 
having many merits and explaining a number of puzzles, led to several 
new puzzles, some of which are listed below.
\begin{itemize}
\item The relevant saddles are critical points of the exact effective 
action, which can typically be thought of as containing fermionic loop 
effects. However according to Picard-Lefschetz theory it is the classical 
action, rather than the quantum one, that is related to the thimble 
decomposition relevant for the resurgent structure.

\item If the quantum corrections are promoted to the classical action 
(i.e. the $g$-suppressed part of the effective action is promoted to 
be of order unity), such saddles no longer play a role in the resurgent 
expansion\footnote{This observation was used in \cite{Serone:2016qog,
Serone:2017nmd} to avoid dealing with non-perturbative contributions 
entirely, by formally treating $g$-suppressed terms as order-unity terms. 
This procedure reshuffles the perturbation theory,  effectively putting 
some coupling dependence into coefficients of the series expansion. 
With this scaling, the quantum tilt becomes classical and
perturbation theory is Borel summable.}.

\item The saddle points have a non-zero imaginary part, which is 
crucial for the resurgent cancellations, as well as to make sense of 
the real physical contributions. However, they appear to violate 
the intersection theorem of the thimbles and their corresponding dual thimbles.

\item Generic quantum mechanical problems, with classically degenerate 
vacua, may still have different curvatures, resulting in quantum 
non-degeneracy \cite{DST}. After one loop effects are taken into account, 
such systems appear to again be saturated by complex saddles of one-loop 
actions. Unlike fermions which can be easily integrated out exactly, the 
bosonic one loop action has infinitely many terms, and it is unclear how 
to set up a systematic treatment of such problems in the same spirit.

\item Finally, this approach had no adequate treatment of the multi-instanton 
corrections of the problems with quantum degenerate minima (e.g. traditional 
double-well and sine-Gordon problems), except that they could be 
treated as a limiting case. While avoiding some of the difficulties 
encountered in the BZJ prescription this is not entirely satisfactory, 
as we expect that such difficulties should possibly be overcome directly 
by a complete Picard-Lefschetz 
theory of path integrals.
\end{itemize}

The main objective of this work is to resolve these discrepancies and 
consolidate the Picard-Lefschetz theory with the resurgence properties
which relate perturbation theory to the saddles of the quantum action.
In particular, we aim to explain the precise relation between critical 
points at infinity associated with the classical action,  their Lefschetz 
thimbles, and the bion solutions to the quantum action\footnote{Indeed, it 
is likely the case that generic critical points of the path integral are of 
this type, including QFT applications.}.
We  show that:
 
\begin{enumerate}[label={\bf \arabic*)}]
\item  Due to their non-Gaussian nature, the integral over the thimble 
attached to the critical point  at infinity may not be saturated in the 
vicinity of the critical point. In fact, it may well be the case that 
this contribution vanishes completely, concealing the effect of the 
saddle on the physical observables. This is in sharp contrast with 
the Gaussian saddles which typically appear in ordinary integrals, 
where the dominant contribution always originates from the saddle itself. 

\item Due to the effects of the Gaussian directions, the main semi-classical 
contribution may arise from the {\it tail} of the QZM descent cycle 
$\Gamma_{\rm QZM}$, which is a consequence of the non-Gaussian nature of 
the  quasi-zero mode directions. In the cases studied here we show that the 
bion saddles of the one-loop (quantum) action (for $p\neq 0$) dominate 
the integration over $\Gamma_{\rm QZM}$. Note that despite the fact 
that the bions are not saddles of the classical action,  and the genuine 
critical point attached to the $\Gamma_{\rm QZM}$ thimble is at infinity, 
it is still the bion configuration that dominates the observables, 
resolving the puzzles outlined above.

\item Such complex configurations, and not the critical points at infinity,
provide the complex phases that lead to hidden topological angles
\cite{Behtash:2015kna}, which are crucial for explaining quantum 
interference effects and QM supersymmetry breaking 
\cite{Behtash:2015loa,Behtash:2015zha}.

\end{enumerate}

 We  argue that these results are generic features of non-Gaussian 
critical points at infinity, and that this structure persists at 
higher orders in the semi-classical cluster expansion. 

 In order to understand the relation between bion amplitudes and correlated 
multi-instanton amplitudes, we also discuss the bion fluctuation operators, 
which determine the leading pre-exponential factor of the corresponding 
amplitudes. The fluctuation operators for bion solutions are quite different 
from the familiar fluctuation operators that appear in the context of 
instantons  \cite{Coleman:1978ae,ZinnJustin:2002ru}. The typical 
fluctuation operator encountered in instanton problems is a Schr\"odinger 
operator with a single-well potential, for example the P\"oschl-Teller 
potential as in \eqref{eq:sg-ifluc} below. For real bions, the fluctuation 
operator involves a double-P\"oschl-Teller potential, see 
\eqref{eq:sg-real-bion-fluctuation} and Fig.~\ref{fig:sg-real-bion-fluc}. 
For complex bions, the fluctuation operator contains even more exotic 
{\it complex} potentials. (For periodic bosonic potentials, the complex bion 
fluctuation potentials can even be singular \cite{Behtash:2015loa,Behtash:2015zha}.) For 
both real and complex bions, the associated fluctuation operators possess 
parametrically small eigenvalues that correspond to non-Gaussian directions.  
When this soft mode is carefully treated, there is perfect agreement  
between the results of the instanton gas analysis and the bion analysis. 

 Our work builds on earlier studies devoted to the role of complex saddle 
points, for example Brezin et al.~\cite{Brezin:1976wa}, Balian, Parisi 
and Voros \cite{Balian:1978et}, Richard and Rouet \cite{Richard:1981gn},  
Lapedes and Mottola \cite{Lapedes:1981tz}, Millard \cite{Millard:1984qt}, 
and Balitsky and Yung \cite{Balitsky:1985in}. Complexified classical 
solutions and Lefschetz thimbles have also been very recently studied 
by Nekrasov using techniques and ideas from the Bethe/gauge correspondence 
\cite{Nekrasov:2018pqq}.

\section{Classical action and critical points at infinity}

 Consider the classical  bosonic Euclidean path integral 
\begin{align}
&Z_{\rm bos} =   \int Dx(t) \;  e^{-\frac{1}{g} S[x(t)]}  
 = \int Dx(t) \;   
    e^{ - \frac{1}{g}  \int dt  \left( \half \dot x^2 + V_{\rm bos} (x) \right)}\, ,  \cr
 &V_{\rm bos} (x)=  \half (W'(x))^2   \, .  
\label{bos-1}
\end{align}
There are instanton and anti-instanton saddle points, which are (real)
solutions of   
\begin{align}
 \dot x =  \pm  W'(x) \, . 
\label{inst-1}
\end{align}
In this non-perturbative normalization, the first order BPS equation and 
its solution are independent of the coupling constant $g$. We concentrate on the 
Sine-Gordon (SG) system, but a similar analysis can be done for the double-well 
system extensively studied in \cite{Behtash:2015loa,Behtash:2015zha}. The 
superpotential and the instanton solution are, respectively, given by 
\begin{eqnarray}
W(x)=4 \cos \left(\frac{x}{2}\right) \quad \Rightarrow \quad  
x_I(t)= 4\arctan\left(\exp[ t-t_{c}]\right)\, , 
\label{eq:sg-instanton}
\end{eqnarray}
and the corresponding instanton action is
\begin{eqnarray}
S_I=\frac{8}{g}\, . 
\label{eq:sg-si}
\end{eqnarray}
Here, $t_c \in \R $ is the position modulus, a bosonic zero mode of the 
instanton solution. The instanton amplitude is given by 
\begin{eqnarray}
\I \equiv  \xi=  J_{t_c}  \; e^{-S_I} 
       \;  \left[\frac{\hat{\det}\, \mathcal{M}_I }
                      {\det \mathcal{M}_0}\right]^{- {1 \over2}}   
  P_I(g)\, ,
\label{instanton-amp}
\end{eqnarray} 
where $J_{t_c} =\sqrt{S_I/(2\pi)}$ is the Jacobian associated with the 
bosonic zero mode, and $\mathcal{M}_I =-\frac{d^2}{d t^2}+V''(x)|_{x=x_{I}(t)}$
is the quadratic fluctuation operator in the background of the instanton.
It has the familiar P\"oschl-Teller form
\begin{eqnarray}
\mathcal{M}_I&=&-\frac{d^2}{d t^2} +1 -2\, {\rm sech}^2(t-t_c) \, , 
\label{eq:sg-ifluc}
\end{eqnarray}
whose only exact zero mode is given by $\dot{x}_I(t)$. The ``hat" on 
${\hat{\det}\,{\mathcal{M}}_I }$ 
indicates that the zero mode has to be removed, and $\det \mathcal{M}_0$ 
is a normalization factor, which we take to be the corresponding free 
fluctuation operator. $P_I(g)$ denotes the perturbative expansion around 
the instanton. 

The determinant of the instanton fluctuation operator can be computed 
using the Gel'fand-Yaglom (GY) method \cite{Gelfand:1959nq,Coleman:1978ae,
Coleman:1977py,Callan:1977pt,ZinnJustin:2002ru,Kirsten:2001wz,Kleinert:2004ev,Dunne:2007rt,Marino:2015yie},
summarized in  Appendix \ref{app}.
For the instanton in the SG model the determinant ratio is a pure number
\begin{eqnarray} 
\frac{\hat{\det}\,\mathcal{M}_I}{\det \mathcal{M}_0}  &=&   \frac{1}{4} \, . 
\label{eq:sg-idet}
\end{eqnarray}
In Section \ref{sec:bions} we will also need the fluctuation determinants 
about more general saddle configurations such as the real and complex bions.
These are quite different from the instanton fluctuation determinants, and 
are computed in  Appendix \ref{app}, using a variant of the GY method  
\cite{McKane:1995vp,Kirsten:2001wz,Dunne:2005rt,Dunne:2007rt}. The 
resulting determinant, with zero mode removed, can be simply expressed 
in terms of the action of the saddle configuration and the asymptotic 
behavior of the associated zero mode. 

\subsection{Cluster expansion}

The partition function can be expressed as a {\it cluster (virial) expansion} 
for an interacting gas of instantons \cite{Coleman:1978ae,ZinnJustin:2002ru}.
 Assuming that there is a gap between 
the ground state and the first excited state, in the $\beta \rightarrow \infty$ 
limit,  we can write $Z$ as 
\begin{align}
Z= e^{-\beta E_0 P_0(g)  } \left( 1+ \frac{\xi}{1!}  \int d\tau_1   
  +  \frac{\xi^2}{2!} \int d\tau_1 d\tau_2   \;  e^{- V_{12} }  
  +  \frac{\xi^3}{3!} \int d\tau_1 d\tau_2 d\tau_3   \;   e^{- V_{123} }  
  +  \ldots \right) \, . 
\label{virial}
\end{align}
In these integrals, $\tau_i$ denotes the position of $i$th instanton event, 
and $V_{ijk\dots}$ denotes a many-body interaction. Neglecting the 
interaction between instantons, the sum exponentiates and we obtain the 
dilute instanton gas correction to the vacuum energy  as
\begin{align}
Z_{\rm dilute} &= e^{-\beta E_0 P_0(g)  } \left(  \sum_{n=0}^{\infty} \frac{\xi^n}{n!}  
       \int d\tau_1 \ldots d\tau_n   \right)   \cr
 & =e^{-\beta E_0 P_0(g)  } \left(  \sum_{n=0}^{\infty} 
         \frac{\xi^n (\int d\tau_1)^n}{n!}  \right)
   = e^{-\beta E_0 P_0(g)}   e^{\beta \xi}   =  e^{-\beta (E_0 P_0(g) - \xi)} \, .
\label{bos-inst-2}
\end{align} 
Note that the instanton-induced term provides 
a {\it negative} contribution to the ground state energy. In fact, this 
argument suffices to show  that any real saddle contribution to the ground 
state energy, in the absence of a $\theta$-angle or Berry phase, is 
universally negative. As was noted in \cite{Behtash:2015kva, Behtash:2015loa,Behtash:2015zha}, 
this observation is at odds with supersymmetry.

 The  interaction terms  $V_{12}, V_{123}, \ldots $ are  functions of the 
separation between the instantons. The two-body interaction can be written
as  $V_{12}= V_{12} (\tau_1 -\tau_2) $ and the three-body term is $V_{123}:=  
V_{12} (\tau_1 -\tau_2) +  V_{23} (\tau_2 -\tau_3) +  V_{31} (\tau_3 -\tau_1)$, 
where the genuine three-body interactions were neglected.  Hence, using the 
relative coordinates $\tau_{ij}= \tau_i - \tau_j$ in each term 
of \eqref{virial}, one of the integrals yields a factor of $\beta$:
\begin{align}
Z= e^{-\beta E_0 P_0(g)  } \left( 1+ \frac{\xi}{1!} \beta 
  + \frac{\xi^2}{2!}  \beta \int d\tau_{12}   \;  e^{- V_{12} (\tau_{12}) }  
  +  \frac{\xi^3}{3!}  \beta \int d\tau_{12} \;  d\tau_{23}   \;   e^{- V_{123} }  
  + \ldots \right) \, . 
\label{virial-2}
\end{align} 
The physical meaning of this factorization is the following: Each instanton 
has one exact zero mode, related to its position. When we have $n$ instantons,  
due to the interaction between them, $(n-1)$ of these zero modes turn into 
quasi-zero modes (QZM), while one of them remains an exact zero mode, related
to the center of mass. The integration over the exact zero mode gives a factor 
of $\beta$. On the other hand, the remaining $(n-1)$-dimensional integral is 
structurally very interesting, and is at the heart of the instanton expansion. 
In particular, the integral $\beta \int d^{n-1} \tau$ behaves as a polynomial 
in $\beta$ with coefficients depending on the coupling constant $g$. It has 
the form   $\beta^{n}(\#) + \beta^{n-1}(\#) + \ldots + \beta^1 (\#)$ that 
contains maximally extensive and sub-extensive terms in $\beta$. This 
structure is indicative of the presence of a critical point at infinity.  

\subsection{Critical points at infinity}

At  finite separation,  there is an interaction between two instantons  
of  the form  
\begin{equation}
 V_{12}(\tau)  = \pm \frac{A }{g} e^{-\tau} \qquad \tau\equiv \tau_{12}, 
\;\;\; A=32, 
\label{int}
\end{equation}
where  $\tau$ is the QZM direction, and $A=32$ with our normalization convention. 
The interaction is repulsive for an
instanton-instanton pair, and attractive for an instanton/anti-instanton 
pair. As a result there is no exact saddle point at any finite separation 
for an $[{\mathcal I}\bar{\mathcal I}]$ pair in the instanton gas picture. 
However, at $\tau=\infty$ there is no interaction between pairs of 
pseudo-particles, and the configuration is indeed a critical point. Below, 
we show that the second order terms in the semi-classical cluster expansion 
generate, for an $[{\mathcal I}\bar{\mathcal I}]$ pair, the contribution
\begin{align}
 \frac{\xi^2}{2!} \int_{\Gamma} d\tau_1 d\tau_2   \;  e^{- V_{12} } 
=  \frac{\beta^2 [{\mathcal I}] [\bar{\mathcal I}] } {2!}
  +   \frac{\beta ([{\mathcal I}\bar{\mathcal I}]_{\pm})^1 }{1!} \, . 
\end{align}
Here, we have performed the integral over the QZM-thimble of the critical 
point at infinity. The first part of this expression (maximally extensive in 
$\beta$) is of course  the uncorrelated (non-interacting) dilute instanton 
gas contribution, and the subextensive $O(\beta)$ term is the leading term 
in the correlated $[{\mathcal I}\bar{\mathcal I}]_{\pm}$ contribution. In 
the cluster expansion, the terms having a form of ${\beta^2 [{\mathcal I}] 
[\bar{\mathcal I}] }/{2!}$ exponentiate and give a contribution of order  
$\beta e^{-S_I}$  to $\beta E_0$, while the summation and exponentiation 
of the terms of the type ${\beta ([{\mathcal I}\bar{\mathcal I}]_{\pm})^1 }
/{1!}$  contribute at order $\beta e^{-2S_I}$ to  $\beta E_0$. The critical 
point at infinity and its QZM-thimble captures both types of contributions. 

\subsection{Overview of the BZJ method} 

 We begin by reviewing the analysis of Bogomolny and Zinn-Justin 
\cite{Bogomolny:1980ur,ZinnJustin:1981dx,ZinnJustin:2002ru}. Their
result is clearly correct -- as it has been checked against numerical 
results and the WKB method -- but it raises some conceptual and 
computational questions that we will resolve below using recent 
advances in resurgence and Picard-Lefschetz theory.  

The second order term in the cluster expansion of the partition function, 
coming from an $[{\mathcal I}\bar{\mathcal I}]$ pair, is
\be
Z_2=  \xi^2 \int_0^\infty d\tau\; e^{\frac{A}{g}e^{- \tau}}\; .
\ee
The above integral is divergent because the integrand in the upper limit 
does not go to zero. This is a benign divergence, which appears in any 
virial expansion. The solution is that we have to subtract the uncorrelated
term, which exponentiates to the dilute instanton gas result, and then 
proceed to compute the correlated remainder,  being a genuine 
instanton/anti-instanton effect. We write ${\rm Virial}_2$, using the 
``-1+1 trick", see e.g. \cite{friedli_velenik_2017}:
\be
Z_2= \xi^2  \int_0^\infty d\tau\; 
  \left( e^{\frac{A}{g}e^{-\tau}}  -1 +1\right) \; .
\ee
The ``+1'' term leads to the appearance of  $\frac{\xi^2}{2!} (\int d\tau_1)^2$ in \eqref{virial}, 
and accounts for the leading order semi-classical expansion which captures 
the effect of non-interacting instantons in the dilute instanton gas 
approximation. The correlated instanton/anti-instanton amplitude 
is
\be
[{\mathcal I}\bar{\mathcal I}] = \xi^2 \int_0^\infty d\tau\;
        \left(e^{\frac{A}{g}e^{-\tau}}-1\right)\;.
\label{IIbar}
\ee
This integral is convergent, but it is dominated by the regime $\tau
\rightarrow 0$ where the notion of an instanton/anti-instanton pair 
does not make sense. 

 To evaluate this integral Bogomolny and Zinn-Justin proposed to analytically 
continue the coupling constant as $g\rightarrow -g$ \cite{Bogomolny:1980ur,
ZinnJustin:1981dx,ZinnJustin:2002ru}. By doing so one obtains a repulsive 
potential between the instanton and anti-instanton. Next, upon integrating 
by parts, the integral reduces to
\begin{align}
[{\mathcal I}\bar{\mathcal I}] & \rightarrow 
       - \frac{A}{g}   \int_0^\infty d\tau\,
     e^{- \frac{A}{g}e^{-\tau}}   e^{-\tau} \tau   \; \cr
&= -\gamma -\log \left(\frac{A}{g}\right)-\Gamma \left(0,\frac{A}{g}\right) \cr
& = -\gamma -\log \left(\frac{A}{g}\right)- e^{-A/g} \left(\frac{g}{A}
    + O\left(g^2\right)\right) \, . 
    \label{eq:bzj1}
\end{align}
BZJ drop the exponentially small terms in this expression, coming from 
$\Gamma \left(0,\frac{A}{g}\right)$, and then continue back to the positive 
coupling constant,  $-g \rightarrow + g$, where one obtains the result
\be
\label{eq:BZJ}
[{\mathcal I}\bar{\mathcal I}]_{\pm}= \mp i\pi \, -\gamma -\log \left(\frac{A}{g}\right)  +\dots
  \, . 
\ee
Here, the sign ambiguity of the imaginary part depends on whether one 
analytically continues back in the upper or lower complex $g$ half-plane, 
respectively.  As is well-known \cite{Bogomolny:1980ur,ZinnJustin:1981dx, 
ZinnJustin:2002ru}, this ambiguity in the result of the 
correlated $[{\mathcal I}\bar{\mathcal I}]_{\pm}$ amplitude cancels exactly 
the ambiguity  that arises by resuming the perturbation theory, which, 
likewise, needs to be defined by analytically continuing the coupling constant
into the complex plane. 

 A problem with this analysis is that it is not clear why it is possible 
to drop terms in \eqref{eq:bzj1} which are exponentially small when $\Re(g)
<0$, but which become exponentially large $\sim e^{+A/g}$  when  analytically 
continued back to $\Re(g)>0$. In the next section we show that if one properly 
treats critical points at infinity via the Lefschetz  thimble decomposition, 
such issues are resolved.

\subsection{QZM-thimble} 
 
The thimble approach to the cluster expansion, based on Picard-Lefschetz 
theory, provides a conceptually cleaner and more systematic way of computing 
multi-instanton contributions. We consider the quantum mechanical system 
with periodic boundary conditions on a temporal circle of length $\beta$. 
The amplitude for the correlated instanton-anti-instanton pair is  
\be
[{\mathcal I}\bar{\mathcal I}] =\frac{1}{2}\int_0^\beta d\tau\; 
e^{\frac{A}{g}\left(e^{-\tau}+e^{-(\beta-\tau)}\right)}-\beta/2\; .
\label{QZM-1}
\ee
The modified instanton interaction reflects the periodicity of the temporal 
box \footnote{More precisely, the exponent leading to the interaction between 
the instanton and anti-instanton must be periodic under  shift symmetry, 
$\tau \rightarrow \tau + \beta $, and is thus given by $ \frac{A}{g}  
\sum_{n=-\infty}^{\infty}  e^{- |\tau + n \beta |}.$ Except for the $n=0,-1$ 
terms that are indeed captured in \eqref{QZM-1}, the rest gives sub-leading 
contributions in $e^{-\beta}$ and are henceforth ignored.}.
Notice that since the above integral counts the $[{\mathcal I}
\bar{\mathcal I}]$ configuration twice (that is, integration from $0$ 
to $\beta/2$ is the same as from $\beta/2$ to $\beta$), we have included 
an extra factor $1/2$.

\begin{figure}[t]
\centering
\includegraphics[scale=0.35]{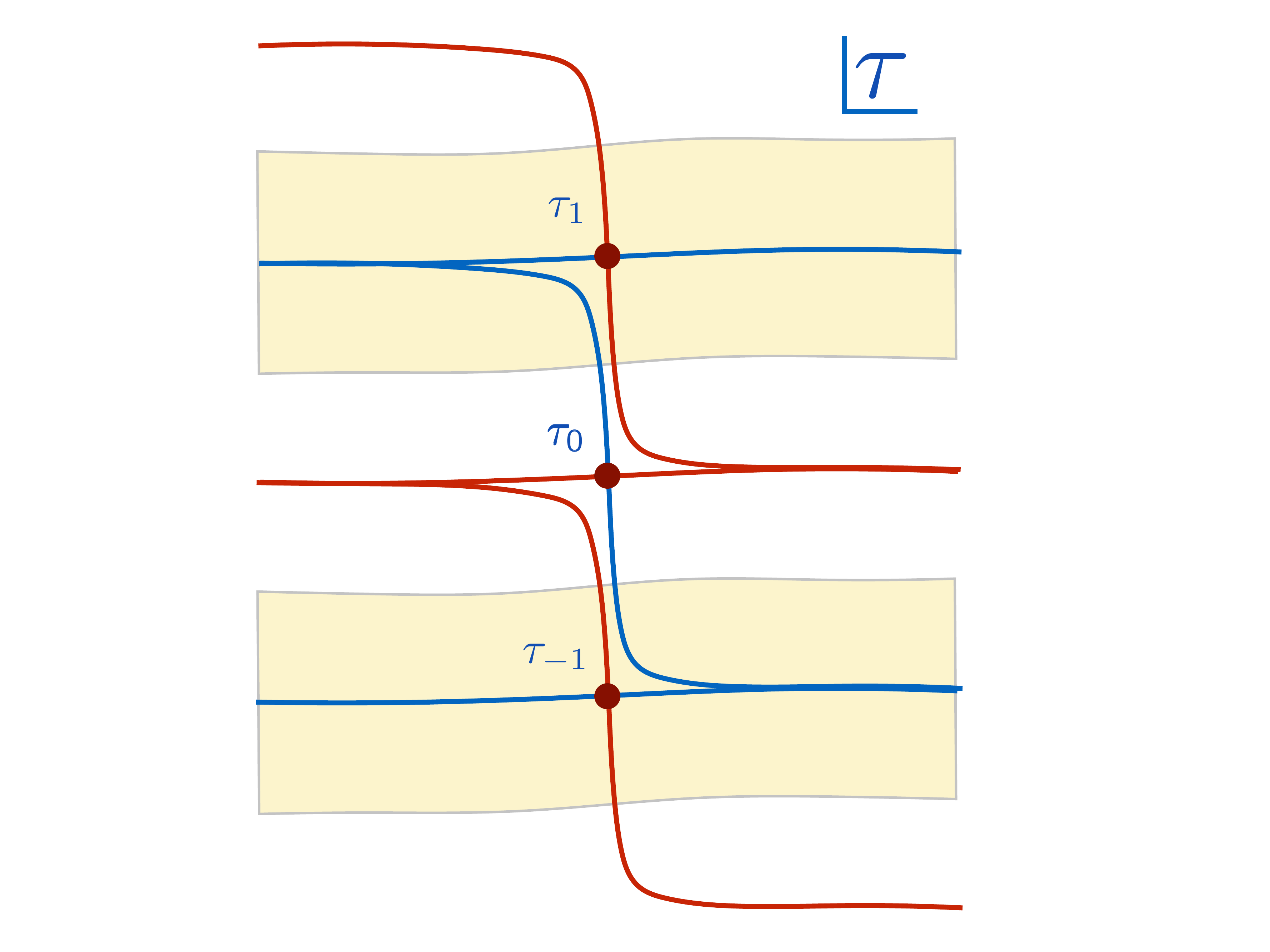}
\caption{The Lefschetz thimbles for the ${\mathcal I}\bar{\mathcal I}$ saddle, 
showing the downward flows (blue curves) connecting $\tau_0$ to $\tau_{\pm 1}$
when $g\to g\, e^{i\theta}$ with $\theta\to 0^+$. The red curves are the 
corresponding upward flows. The directions are
flipped about the imaginary axis for $\theta\to 0^-$.}
\label{fig:thimbles}
\end{figure}

The exponent of the integrand in \eqref{QZM-1}  has a critical point, which 
corresponds to an \emph{exact instanton/anti-instanton solution} 
\cite{Lapedes:1981tz}. The precise form of this solution is not relevant 
for our argument\footnote{Such solutions obviously exist, however, whenever 
there are two neighboring classical minima. To see this, consider a particle 
moving in the inverted classical potential (e.g. inverted double-well). We 
seek a solution which interpolates from the vicinity of one classical vacuum 
to another (i.e. one peak of the inverted potential to the other). These 
solutions exist and show an oscillatory behavior between the two peaks, 
with a period exactly equal to $\beta$.}. The 
$[{\mathcal I}\bar{\mathcal I}]$ action
\be
S(\tau)=-\frac{A}{g}\left(e^{-\tau}+e^{-(\beta-\tau)}\right)
\ee
correctly captures the action in the QZM direction up to higher order 
terms in $e^{-\tau}$. The critical point of $S(\tau)$ is easily seen 
to be $\tau_0=\beta/2$ even though there are other saddles as well, 
see e.g.~\cite{Lapedes:1981tz}, at $\tau_n=\beta/2+i\pi n$. We now 
transform $\tau=\tau_0 +\xi$. However, it is immediately observed that 
along the original integration cycle (i.e. $\R$) the action decreases 
away from the critical point $\xi=0$. This just reflects the fact that 
the instanton--anti-instanton pair {\it attracts} for real separations 
$\tau$ (i.e. the critical point at infinity is an ``unstable'' saddle 
in the traditional sense). The Picard-Lefschetz theory, however, the 
contribution of the saddle should be evaluated along the ``downward flow'', 
where the action always increases, thus ensuring convergence. Such 
downward flow emanates from $\tau=\tau_0$ directly into the imaginary 
$\tau$-direction, which is sketched in Fig.~\ref{fig:thimbles}. The 
downward flow from the saddle at $\tau_0$ connects to the saddles 
$\tau_{\pm 1}$, meaning that we are on a Stokes line. This scenario 
agrees with what one expects from perturbation theory. To move off the 
Stokes line, we analytically continue $g\rightarrow g e^{i\theta}$. For 
$\theta$ being small and positive, the picture in Fig.~\ref{fig:thimbles} 
is obtained. For $\theta$ small and negative, the picture in 
Fig.~\ref{fig:thimbles} is flipped about the vertical axis.  

Now it is clear that the downward flow, or the Lefschetz thimble, through 
the saddle at $\tau=\tau_0$ consists of three parts; namely
\begin{align}
&\Gamma_{\rm QZM}^{\theta=0^{\pm}}  = \gamma_1^\pm   \cup \gamma_2^\pm 
 \cup  \gamma_3^\pm,
\end{align}
where the segments are
\begin{align}
& \gamma_1^\pm: \xi\in(-\infty\pm i\pi,\pm i\pi), \qquad 
  \gamma_2^\pm: \xi\in[\pm i\pi,\mp i\pi], \qquad 
  \gamma_3^\pm: \xi\in (\mp i\pi,\infty\mp i\pi) \, . 
\end{align}
We therefore define
\be \label{eq:IIbarfinal}
[{\mathcal I}\bar{\mathcal I}]_\pm=\lim_{\beta\rightarrow\infty}
 \frac{1}{2}\left(\int_{\cup_{i}\gamma_i^\pm} d\xi\; 
   e^{-S(\xi)}-\beta \right) \;
\ee
Evaluating the integral along $\gamma_2^\pm$ in the limit $\beta\rightarrow
\infty$ gives $\mp i\pi$. Now, it is left to compute the integral over
$\gamma_1^\pm\cup\gamma_3^\pm$. This can be concisely written as
\begin{multline}
\int_{\gamma_1^\pm\cup\gamma_3^\pm}d\xi\;e^{-S(\xi)}
  = \int_{-\infty}^\infty d\xi\;e^{-\frac{2A}{g}e^{-\beta/2}\cosh(\xi)}
  = 2 K_{0}\left(\frac{2A}{g}e^{-\beta/2}\right)\; ,
\end{multline}
where $K_0(z)$ is a modified Bessel function of the second type. Taking 
$\beta\rightarrow \infty$, we can use the asymptotic behavior of $K_0(z)$
\be
K_0\left(\frac{2A}{g}e^{-\beta /2}\right)
  \rightarrow -\gamma -\log\left(\frac{A}{g}\right)+\beta/2 \, . 
\ee
The amplitude \eqref{eq:IIbarfinal} finally takes the form 
\be
[{\mathcal I}\bar{\mathcal I}]_\pm=\mp i\pi \, -\gamma
  - \log\left(\frac{A}{g}\right) +\dots 
\, . 
\label{eq:IIbar-BZJ}
\ee
Clearly, this construction does not run into the conceptual difficulty 
mentioned in the previous section. It also clarifies that in path integration 
the field space (and specifically the quasi-zero mode direction) must be 
complexified in order to properly define the semi-classical expansion. 
Finally, it shows that the Lefschetz thimble associated with a critical 
point at infinity (not necessarily the critical point {\it per se}), yields 
the dominant contribution. We will see that the final point becomes more 
relevant once fermions are included. In that case, the critical point at 
infinity does not contribute at all, and the entire contribution comes 
from a non-Gaussian thimble. 

\section{Quantum action and bions} 
\label{sec:bions}

 In this section we analyze the role of critical points at infinity
in quantum mechanical models containing fermions. We consider the 
Euclidean path integral  
\begin{align}
Z =   \int Dx(t) \;  e^{-\frac{1}{g} S[x(t)]}  
 = \int Dx(t) \;   
    e^{ - \frac{1}{g}  \int dt  \left( \half \dot x^2 + V(x) \right)}\, , 
\label{R-1}
\end{align}
where the  bosonic potential is (recall that $W(x)=4\cos\left(\frac{x}{2}
\right)$ for the SG system)
\begin{align} 
V(x)=  \half (W'(x))^2  + \tfrac{p\, g}{2}  W''(x).
  \label{Gen-NP-3}  \qquad 
\end{align}
The important feature of the action in \eqref{R-1} is that in addition
to having a purely classical part, it has a quantum piece that is of the form 
\begin{align} 
V(x) = v_0(x)  +  p\, g\,   v_1(x)  \, ,  
\end{align}
where the term $p\,g\,v_1\equiv \tfrac{p\,g\,}{2}  W''(x)$ arises from 
integrating out the fermionic degrees of freedom \footnote{If the second 
term is a classical tilt, i.e. $pg\sim O(1)$, of the potential, then 
complex field configurations still exist, but perturbation theory 
is Borel summable, see \cite{Serone:2017nmd}.}. 
We refer to $V(x)$ as the effective quantum potential.  

 For $p=\pm 1$, the quantum potential corresponds to the bosonic and 
fermionic sectors of supersymmetric QM. For other positive or negative 
integers, the quantum potential is related to  quasi-exactly solvable 
systems. This can be seen by integrating out $N_f$ fermionic fields 
in the action 
\begin{equation}
S = \frac{1}{g}  \int dt \left( \half \dot x^2  + \half (W')^2 
   + \half  (\bar \psi_i \dot \psi_i -   \dot {\bar \psi}_i \psi_i) 
   + \half  W'' [\bar \psi_i, \psi_i ]  \right)\, ,  
   \qquad i=1, \ldots, N_f \, . 
\label{lag}
\end{equation}
The system given in \eqref{R-1} with the potential function \eqref{Gen-NP-3}  
has exact solutions of the equations of motions associated with the quantum 
potential \cite{Behtash:2015loa,Behtash:2015zha}
\begin{align}
  +  \frac{d^2 z}{dt^2}  = \frac{\partial V}{\partial {z}}  
        \qquad    {\rm or}  \qquad 
    \frac{d^2 z}{dt^2}  = W'W'' + \tfrac{p\, g}{2} W''' \, . 
\label{PLW-3}
\end{align}
These exact solutions inlcude the bounce as well as real and complex 
bions. The explicit form of the solutions is summarized in Appendix 
\ref{app}. Unlike instantons, the exact solutions of the equations 
of motion involve $g$, and have a characteristic size ${\rm ln}(A/g)$, 
which can be interpreted as the separation of instantons-anti-instantons
constituents.

It is important to note that the classical solutions, the leading saddle 
points coming from the classical action, are still given by instantons. 
At second order in the cluster expansion, two instantons at infinite 
separation correspond to a genuine critical point at infinity. In this 
section we explain the precise relation between these classical saddle 
points at infinity and the exact bion solutions of the quantum action.

\subsection{Critical points at infinity and their thimbles vs. 
exact bion solutions}
 
Consider the quantum system on a temporal circle with size $\beta$,   
as in the $p=0$ case. The presence of fermions, or equivalently, the existence 
of the quantum term in the potential, modifies the interaction between 
instantons. We denote the interaction potential between two instantons by 
${\cal V}_{+}(z)$, and the potential between an instanton and an anti-instanton 
by ${\cal V}_{-}(z)$:
\begin{align}
\label{interaction}
{\cal V}_{\pm}(\tau) =  \pm \frac{A}{g} 
  \left(e^{- \tau} +e^{- (\beta -\tau)} \right)+   p \, \tau \, . 
\end{align}
The interaction between instantons has both classical and quantum terms. 
The classical bosonic interaction is repulsive for correlated $[{\mathcal I}
{\mathcal I}]$ pairs and attractive for $[{\mathcal I}\bar{\mathcal I}]$ 
pairs. On the other hand, the fermion zero mode (or quantum)  induced 
potential, $\tfrac{p\,g}{2} W''(x)$,  leads to an   attractive interaction for  both $[{\mathcal I}
{\mathcal I}]$ and $[{\mathcal I}\bar{\mathcal I}]$ pairs. 

\begin{figure}[t] %
\centering
\includegraphics[scale=0.5]{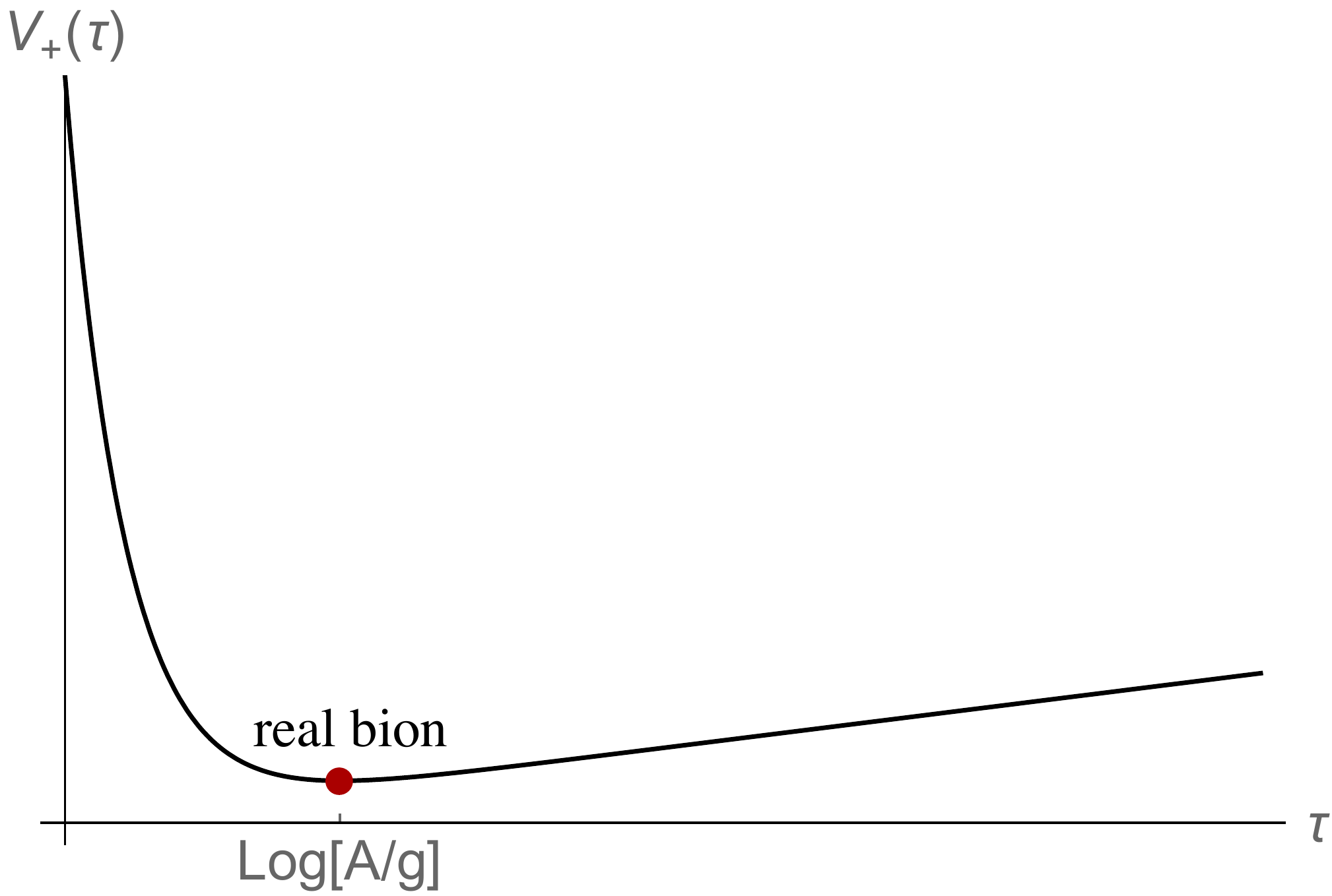}%
\caption{The effective interaction potential $\cal{V}_+(\tau)$ associated 
with the $[{\mathcal I}{\mathcal I}]$  configuration. The critical 
value $\tau^*= \ln \left( \frac {A}{g\,p} \right)$ gives the dominant 
contribution to the $[{\mathcal I}{\mathcal I}]$ amplitude integral 
in \eqref{eq:ii-integral}.}
 \label{fig:quantum-ii-potential}
\end{figure}

\subsubsection{$[{\mathcal I}{\mathcal I}]$  thimble integration}
\label{sec:ii-thimble}

We discuss first the slightly simpler case of an $[\I\I]$ pair configuration. 
This configuration is a critical point at infinity, which means that $\tau_0
=\beta/2$ goes to infinity as  $\beta\to\infty$. In contrast to a Gaussian 
critical point, the contribution from this critical point vanishes as 
$\beta\to\infty$, 
\begin{equation}
\lim_{\beta\rightarrow\infty} e^{-\frac{2A}{g}\left(e^{- \beta/2} \right)}  
  e^{-p  \beta/2}  =0\, . 
\end{equation}
This does not mean that the contribution of the thimble vanishes. Indeed, 
the exact integration over  $\Gamma_{\rm QZM}^{\theta=0^{\pm}} $ in the $\beta 
\rightarrow \infty$ limit gives
\begin{align}
I_+(p, g)\equiv  \int_{\Gamma_{\rm QZM}^{\theta=0^{\pm}}}  d\tau\; 
   e^{-\frac{A}{g}\left(e^{-\tau}+e^{-(\beta-\tau)}\right)} e^{-p  \tau}    
  =   \left(\frac{g}{A}\right)^{p} \Gamma(p)  
  \label{eq:ii-integral}
\end{align}
The $[\I \I]$ amplitude, derived from integrating over the associated 
QZM-thimble, is 
\begin{align}
[\I\I] 
   &=  I_{+}(p, g) \times  [\I]^2  \cr
   &=  \left(\frac{g}{A}\right)^{p} \Gamma(p)   \times 
       \tfrac{S_I}{2\pi} \textstyle{ 
         \left[\frac{\hat{\det}\,\mathcal{M}_I}{\det \mathcal{M}_0}\right]^{-1}}
         e^{-2S_I}  \,   \cr
   & =   \frac{1}{2 \pi}  \left(\frac{g}{32}\right)^{p-1} \Gamma(p)    
        e^{-2S_I}   \,   .
\label{II}
\end{align}
The leading contribution to the integral in \eqref{eq:ii-integral} arises 
not from the bosonic critical point at $\tau_0=\beta/2$, but instead from 
the {\it tail} of the thimble attached to the critical point $\tau^{*} =  \ln 
\left( \frac {A}{g\,p} \right)$. See Fig.~\ref{fig:quantum-ii-potential} 
for a sketch of the form of the $[{\mathcal I}{\mathcal I}]$ interaction 
potential. The scale corresponding to this dominant contribution
\begin{align}
\label{bion-2}  
\tau^{*} =  \ln \left( \frac {A}{g\,p} \right)  \, , 
\end{align}
is interpreted physically as  the separation between the two instantons 
\cite{Behtash:2015zha,Behtash:2015loa}. In Section \ref{sec_rb} we show 
the result \eqref{II}, including the pre-exponential terms, matches
the contribution of the real bion solution of the quantum potential.

\subsubsection{$[{\mathcal I}\bar{\mathcal I}]$  thimble integration}
\label{sec:iibar-thimble}

We now consider the effect of an $[\I\bar \I]$ pair. The integration over 
the quasi-zero mode direction now  becomes
\be
[\I\bar \I]_{\rm naive}= \int_0^\beta d\tau\; 
  e^{\frac{A}{g}\left(e^{-\tau}+e^{-(\beta-\tau)}\right)} e^{-p  \tau} \, . 
\label{QZM-2}
\ee
Here, we have written the integral along the positive real line. This
is clearly too naive. The QZM integration should be carried out 
over the Lefschetz thimble associated with the critical point. The 
critical point at infinity, $\tau_0=\beta/2$, of the bosonic action is the same as for the 
$p=0$ theory. 
The contribution of the vicinity of this critical point  is
\begin{align}
& \int_{\Gamma_{\rm critical \;  point}}  d\tau\; 
     e^{\frac{A}{g}\left(e^{-\tau}+e^{-(\beta-\tau)}\right)}e^{-p  \tau}  
\approx  e^{\frac{2A}{g}\left(e^{- \beta/2} \right)}  e^{-p  \beta/2}  \, , 
\label{QZM-3} 
\end{align}
which vanishes in the limit $\beta \rightarrow \infty$
\begin{align}
&\lim_{\beta\rightarrow\infty} e^{\frac{2A}{g}\left(e^{- \beta/2} \right)}  
   e^{-p  \beta/2}  = 0 \, . 
\end{align}
As for the $[{\mathcal I}{\mathcal I}]$ saddle, this does not mean that the contribution of the thimble vanishes.

\begin{figure}[t]
\centering
\includegraphics[scale=0.55]{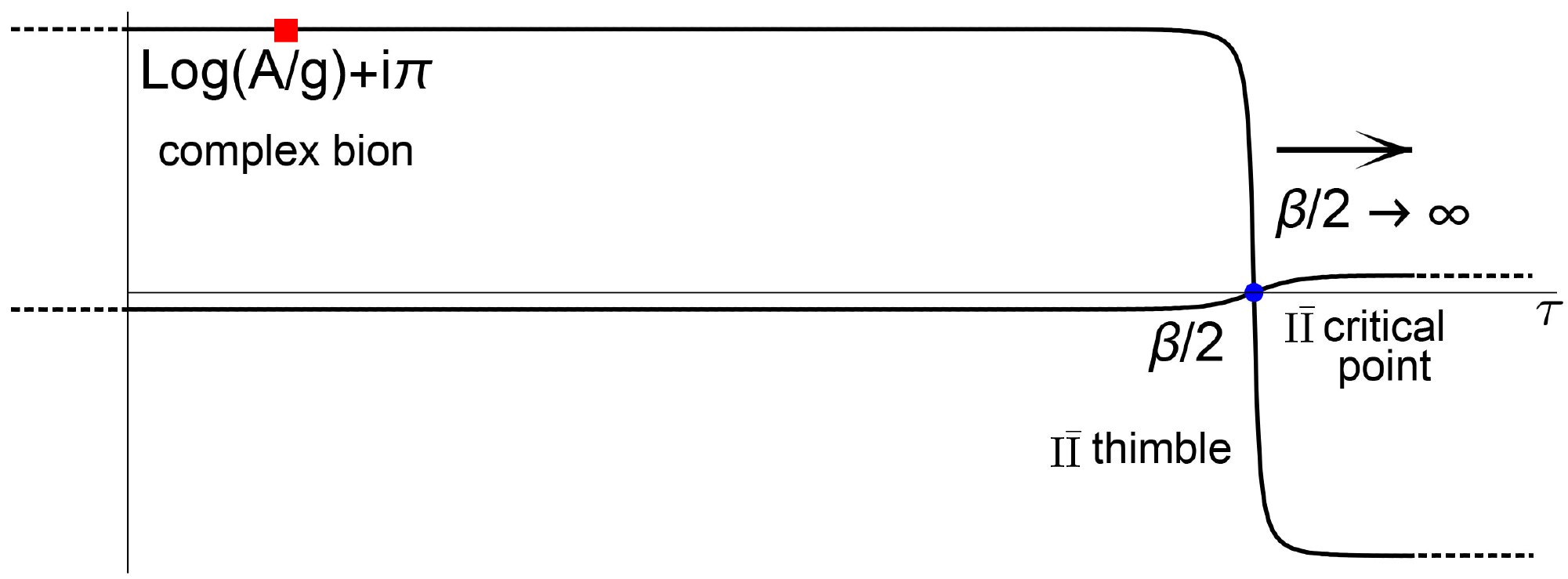}
\caption{
Plot of the Lefschetz thimble attached to the $[{\mathcal I}\bar{\mathcal I}]$ 
critical point at $\tau_0=\beta/2$. As $\beta\rightarrow \infty$, the critical 
point moves to infinity and its contribution to the thimble integral 
vanishes. However, the integral receives a non-vanishing contribution 
from a complex bion configuration located at $\tau^*_+=\log(A/g)+i\pi$. Compare 
with Fig.~\ref{fig:quantum-ii-potential}.}
\label{fig:iibar-thimble}
\end{figure}

 The QZM direction is non-Gaussian, and this fact is encoded in the thimble 
integration. The contour should be deformed along the downward flow direction, 
as in Fig.~\ref{fig:thimbles}. Then the exact integration over the  quasi-zero 
mode thimble $\Gamma_{\rm QZM}^{\theta=0^{\pm}}$ in the $\beta \rightarrow \infty$  
limit results in
\begin{align}
I_-(p, g)\equiv  \int_{\Gamma_{\rm QZM}^{\theta=0^{\pm}}}  d\tau\; 
    e^{\frac{A}{g}\left(e^{-\tau}+e^{-(\beta-\tau)}\right)}e^{-p  \tau}    
 =  e^{\pm i \pi p}  \left(\frac{g}{A}\right)^{p} \Gamma(p)  \, . 
 \label{eq:iibar-integral}
\end{align}
The $[\I\bar \I]$ amplitude, derived from integrating over this QZM-thimble,
is given by
\begin{align}
[\I \bar  \I]_{\pm}  
   &=  I_{-}(p, g) \times  [\I]^2  \cr
   &=  e^{\pm i \pi p}  \left(\frac{g}{A}\right)^{p} \Gamma(p)   \times 
        \tfrac{S_I}{2\pi} \textstyle{ \left[
        \frac{\hat{\det} \,\mathcal{M}_I}{\det \mathcal{M}_0}\right]^{-1}}
        e^{-2S_I} \,   \cr
   & =   \frac{1}{2 \pi}  \left(\frac{g}{32}\right)^{p-1} \Gamma(p)    
        e^{-2S_I}   e^{\pm i \pi p} \,   .
        \label{eq:IIbar}
\end{align}
Here, use was made of the fact that the non-zero mode determinant of 
the $[\I\bar \I]$ approximately factorizes.

This result for the $[{\mathcal I}\bar{\mathcal I}]$ amplitude has a 
number of interesting aspects:
\begin{itemize}
\item Unlike regular Gaussian saddle points for which critical point 
gives the dominant contribution, the critical point at infinity and its 
vicinity do not contribute. This is different from  the usual situation 
in Picard-Lefschetz theory, which by definition has only Gaussian 
modes because, strictly speaking, the ``action'' must be a Morse function. 
\item The integral in \eqref{eq:iibar-integral} is dominated by the 
{\it tail} of the Lefschetz thimble in the sense that the main contribution 
comes from the complex  points
\begin{align}
\label{bion}  
\tau^{*}_\pm = \left[ \ln \left( \frac {A}{g\, p}\right) \pm i \pi \right]\, ,  
\end{align}
where the real part of $\tau^{*}_\pm $ corresponds to the separation between 
the instanton and anti-instanton constituents \cite{Behtash:2015zha,
Behtash:2015loa}. 
\item We  show below in Section \ref{sec:cb} that this contribution corresponds
to the exact complex bion solution of the quantum modified equations of
motion.  
\item By expressing the $[\I\bar\I]_{\pm}$  amplitude as 
\begin{align}
[\I \bar\I]_{\pm} = \frac{1}{2\pi} \left(\frac{g}{32}\right)^{p-1} \Gamma(p)    
      e^{-2S_I}  \left( \cos(  \pi p ) \pm  i \sin(  \pi p ) \right)   \, , 
\end{align}
it is easy to see that the two-fold ambiguous part $\sim \pm  i \sin(\pi p) 
\Gamma(p)e^{-2S_I}$  must be related to resurgent cancellations.  This part 
cancels against the ambiguity in the Borel resummation of perturbation 
theory, as verified in \cite{Behtash:2015zha,Kozcaz:2016wvy}.   

The unambiguous term $\cos(\pi p)$ is the hidden topological angle (HTA) 
associated with the bion configuration. The HTA ensures that the complex 
bion contribution to the ground state energy can be either positive or 
negative depending on $p$, despite the fact that the Lefschetz thimble 
belongs to a {\it real} critical point at infinity \footnote{Recall 
that the contribution of all  real (instanton) configurations to ground 
state energy is negative semi-definite (in the absence of topological 
theta-angles or Berry phases accompanied with the instanton amplitudes.).  
The fact that the complex bion contribution may be positive has very 
important implications both in QM as well as QFT \cite{Behtash:2015kna,
Dunne:2016jsr}. }. In particular, in supersymmetric QM the ground state 
energy vanishes due to an exact cancellation between the semi-classical 
contributions of real and complex bions \cite{Behtash:2015zha,Behtash:2015loa}. 
\end{itemize}

\begin{figure}[t] %
\centering
\includegraphics[scale=0.5]{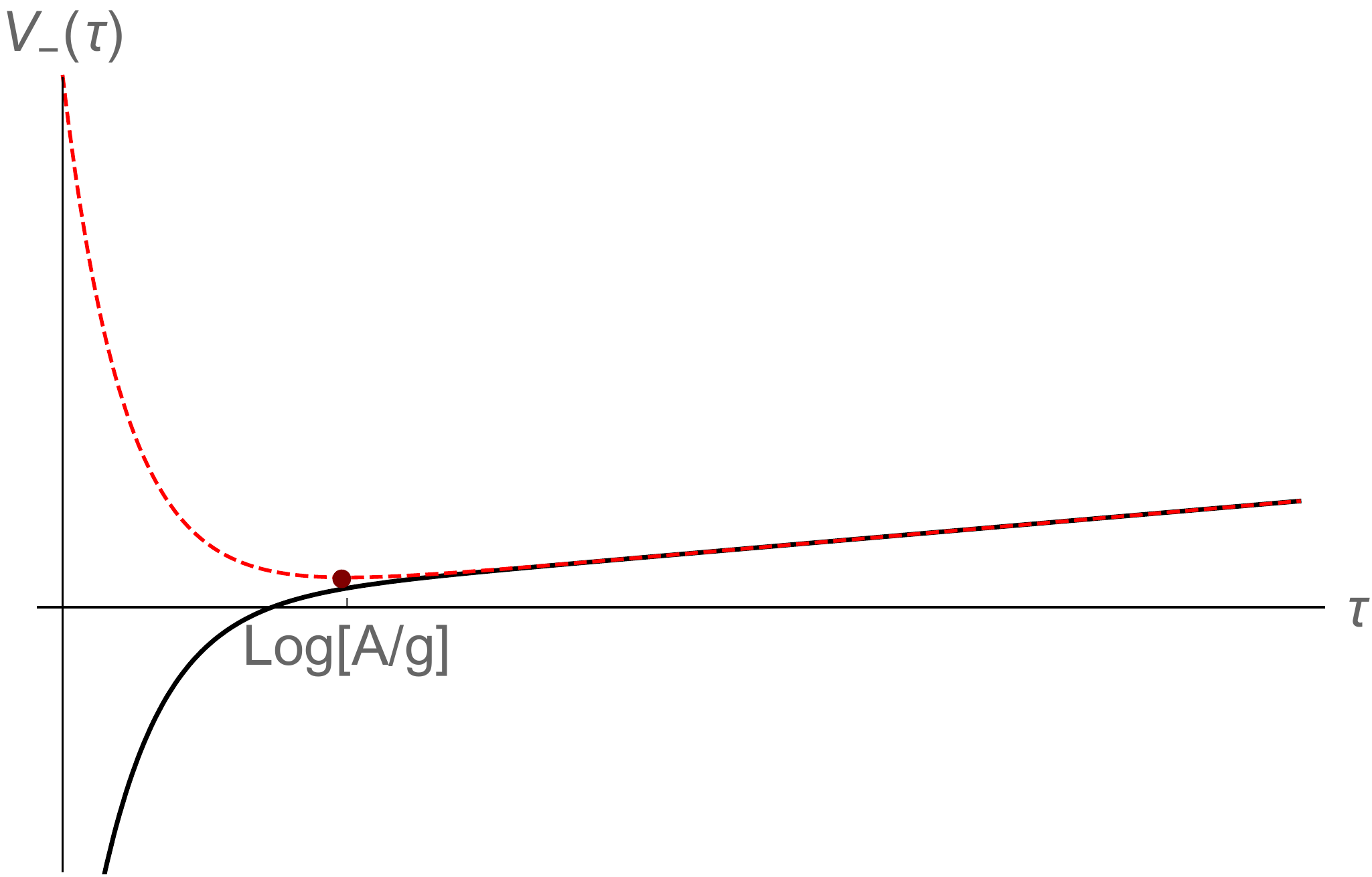}%
\caption{The effective interaction potential $\cal{V}_{-}(\tau)$ associated with 
the $[{\mathcal I}\bar {\mathcal I}]$  configuration.  For real values of 
the separation $\tau$, (black-solid curve),  this completely attractive 
potential  is the reason why the $[{\mathcal I}\bar {\mathcal I}]$ 
configuration is viewed as unstable in the literature. The red-dashed 
curve correspond to the effective potential on the thimble. The value 
$\tau^*= \ln \left( \frac {A}{g\,p} \right)+ i \pi$ gives the dominant 
contribution to the $[{\mathcal I} \bar {\mathcal I}]$ amplitude integral 
in \eqref{eq:iibar-integral}.} 
 \label{fig:quantum-iibar-potential}
\end{figure}

In Fig.~\ref{fig:quantum-iibar-potential} we plot the potential $\cal{V}_{-}
(\tau)$ for $i)$ real values of $\tau$ and $ii)$ for values of $\tau\in
\Gamma_{\rm QZM}$, on the $[{\mathcal I}\bar{\mathcal I}]$ QZM thimble. 
For simplicity, consider the $\beta \rightarrow \infty$ limit. For real 
values, as discussed in textbooks and reviews, the potential (shown by 
the black solid curve) is always attractive and the $[{\mathcal I}
\bar{\mathcal I}]$ configuration is unstable.  There is no length scale 
at which the pair stabilizes. This is the perspective that one obtains 
by using standard techniques such as real gradient flow or the valley 
method.  On the other hand, on the QZM-thimble,  the exponential 
(classical) part becomes repulsive as shown in the red dashed curve. 
The one-loop quantum-induced term leads to an attractive ``force", 
resulting in a stable configuration at $\tau^*= \ln \left(\frac {A}{g\,p}
\right)+ i \pi$. However, since this latter term is quantum mechanical 
in origin, the  minima are not  critical points, but points on the tail 
of genuine $[{\mathcal I}\bar{\mathcal I}]$-thimbles  which dominate 
the integral.

 One may think that this description is essentially the analytic 
continuation of BZJ, $g \rightarrow -g$, which turns the attractive 
potential to a repulsive one. However, this is not true. In the BZJ 
prescription, after analytic continuation,  the range of integration 
for $\tau$ is $[0, \infty)$. This range differs from the one appearing 
on  $\Gamma_{\rm QZM}$. As explained around \eqref{eq:bzj1}, the integration 
on the BZJ domain induces an exponentially small term  $e^{-A/g} $ that  
would become exponentially large $e^{+A/g} $  once one moves back to the 
physical theory. This term was dropped in the BZJ analysis without 
justification. On the other hand, the integration on $\Gamma_{\rm QZM}$ 
does not generate this pathological term.  As emphasized earlier,  
in order to obtain this result, one needs to use the complex gradient 
equation to determine the integration cycle as opposed to the real gradient 
flow or valley techniques.  Once this is done, the analysis of the 
$[{\mathcal I} \bar {\mathcal I}]$ is  fundamentally on the same footing 
as $[{\mathcal I}  {\mathcal I}]$.

\subsection{Real bion amplitude vs. $[\I \I]$ amplitude}
\label{sec_rb}

 The semi-classical treatment of the supersymmetric SG system was shown 
to require the inclusion of the real bion, which is a solution of the 
one-loop quantum action  \cite{Behtash:2015zha,Behtash:2015loa}. The 
properties of the real bion are reviewed in Appendix \ref{sec:det-real-bion}. 
The real bion configuration is a solution of the Euclidean equations of 
motion \eqref{PLW-3} which includes quantum one-loop effects (see 
\eqref{Gen-NP-3}). In this section we explain the relation between 
this solution and the $[\I\I]$ amplitude \eqref{II} computed in Section 
\ref{sec:ii-thimble},  obtained by using the classical action and 
integrating over $\Gamma_{\rm QZM}$. 

 This real bion solution  has 
an exact translational zero mode, the center position of the bion.  It also 
has a parametrically small ``soft mode" that will be important below. The  
amplitude is given by 
\begin{align}
I_{\rm real\, bion} = [{\cal RB}]  & =  J_{t_c}  \;   
   \left[\frac{\hat{\det} \,\mathcal{M}_{\rm real\, bion}}
     {\det \mathcal{M}_0}\right]^{-{1\over2}}
     e^{-S_{\rm real\, bion}}\, . 
\label{amp_rb}
\end{align}
Here,  $J_{t_c}=[S_{\rm real\, bion}/(2 \pi)]^{1/2}$ is the Jacobian 
associated with the translational zero mode of the bion, $S_{\rm real\, bion}$ 
is the real bion action, and the determinant is that of the fluctuation 
operator in the real bion background. We now show that, taking into account 
the soft mode of the fluctuation operator, the real bion amplitude coincides 
precisely with the correlated $[\I \I]$ amplitude in \eqref{II}. This is a 
non-trivial result, because the details of the two calculations differ in 
several significant ways:

\begin{itemize}
\item In the $[\I\I]$ calculation there are two powers of the Jacobian 
$J_{t_c}$, one for each instanton. In the $[\cal RB]$ analysis, however,
there is only one zero mode and the Jacobian enters only once. Note that
the Jacobian is proportional to the square root of the classical action.
Since $S_{\rm real\, bion}\sim 2S_{I}$, it is certainly not the case that the 
Jacobian factors match.  

\item For $[\I\I]$ there is an integral over the 
QZM direction, corresponding to the $[\I\I]$ separation. The size of the 
real bion, on the other hand, is fixed by the value of the coupling constant 
$g$. The existence of a non-Gaussian mode must be encoded in the presence of 
a parametrically small eigenvalue in the fluctuation operator. It is not 
immediately obvious that this contribution matches the one arising from  
the QZM integration.

\item  The two calculations also differ in the treatment of the 
classical $[\I\I]$ interaction. In the conventional instanton calculus, the 
interaction is determined from an ansatz (or an approximate flow equation), 
and then included in the QZM integral. In the $[\cal RB]$ analysis  
\cite{Behtash:2015zha,Behtash:2015loa} we 
observe that the saddle point has action $S_{\rm real\, bion}
 =  2S_I - p\, \log\left( {\frac{p\,g}{32 e}} 
\right) +  \ldots $. 
The logarithmic correction term must combine with the parametrically small 
eigenvalue of the fluctuation operator to build the integral over the QZM 
in the $[\I\I]$ analysis. We now demonstrate this explicitly.
\end{itemize}

The real bion fluctuation determinant is computed in Appendix 
\ref{sec:det-real-bion}. We  combine the Jacobian factor, proportional 
to $(S_{\rm saddle})^{1/2}$,  with the inverse square root of the determinant 
\eqref{eq:sg-real-bion-det} of the fluctuation operator, to obtain the 
semi-classical pre-factor:
\begin{eqnarray} 
    \sqrt{ \tfrac{S_{\rm real\, bion}}{2 \pi}  }     
     \left[\frac{\hat{\det}\mathcal{M}_{\rm real\, bion}}
                  {\det\mathcal{M}_0}\right]^{- {1\over2}}
  &\approx& \frac{1}{2\pi} \left(\frac{32}{g}\right) \sqrt{\frac{2\pi}{p}} \, .  
\label{rb_jac_det}
\end{eqnarray}
Note that in the weak coupling limit, the relation between the 
determinant in the real bion \eqref{eq:sg-real-bion-det} and the instanton 
background \eqref{eq:sg-idet} is given by 
\begin{align}
  \frac{\hat{\det} \mathcal{M}_{\rm real\, bion} }{\det \mathcal{M}_0}
  \approx \frac{pg}{64} =  \frac{pg}{4} \left[\frac{ \hat{\det} \mathcal{M}_{I} }
                                   { \det \mathcal{M}_0}\right] ^2 
  =  \frac{pg}{4} \left[ \frac{ 1}{4}\right] ^2.
\end{align}
This result proves the existence of one parametrically small 
eigenvalue in the real bion background. We can view the presence of 
a small eigenvalue as the result of tunneling between the two minima
in the potential of the fluctuation operator, see  
Fig.~\ref{fig:sg-real-bion-fluc}.  
Expressing the bion action in terms of the instanton action, we get 
\begin{align}
e^{-S_{\rm real\, bion}  }  &\sim \left( {\frac{g}{32}} \right)^{+p} 
    \left( {\frac{p}{e}} \right)^{+p} e^{-2S_I  } \,.
\label{rb_S_cl}
\end{align}
Combining \eqref{rb_jac_det} and  \eqref{rb_S_cl} the real bion amplitude 
is obtained as 
\begin{align}
  [{\cal RB}] = I_{\rm real\, bion} 
   = \frac{1}{2\pi}   \left( {\frac{g}{32}} \right)^{+p-1}   
    \left( \frac{p}{e} \right)^{+p}    
    \sqrt {\frac {2\pi}{p}} e^{-2S_{I}} \, . 
    \label{realbion}
\end{align}
Now compare this real bion result with the $[\I \I]$ amplitude in  \eqref{II}. We 
observe that  $(p/e)^{p} \sqrt {2\pi/p} $ is the asymptotic expansion of 
$\Gamma(p)$ for large $p$, so that the two results agree if we substitute the 
full expression for its asymptotic form. It is instructive to understand why 
\eqref{realbion} only reproduces the asymptotic form. The crucial observation 
is that we have  treated the soft mode in the real bion analysis as if it 
is a Gaussian mode.  

 More specifically, note that the QZM integration (in the $\beta\to
\infty$ limit) is given by 
\begin{align}
 \int_{\Gamma_{\rm QZM}, \rm exact }  d\tau\; e^{- \left( \frac{A}{g} e^{-\tau }  
  + p  \tau  \right) }  
 =    \left(\frac{g}{A}\right)^{p} \Gamma(p)  \, . 
 \label{exact}
\end{align}
If we momentarily ignore the fact that the second term in the exponent 
is a quantum correction, and perform a saddle-point analysis of the 
effective potential $V(\tau)= \frac{A}{g} e^{-\tau } + p  \tau$, we 
find that the critical point is at $ \tau^{*} = \ln(\frac{A}{g\,p})$,
which is the size of the real bion. Using a Gaussian approximation, we find
\begin{align}
 e^{-V(\tau_*)}  \int d \tau   \; e^{- \frac{p}{2} \tau^2}   
    = \left(\frac {p\,g}{Ae}\right)^{p} \sqrt{ \frac {2\pi}{p}}\, ,
\end{align}
which is the leading asymptotic approximation to the exact result given in 
\eqref{exact}. This implies that it is in general not justified to
treat the soft mode in the real bion fluctuation operator in the
Gaussian approximation. If the soft mode of the real bion was treated 
consistently, the result would agree with \eqref{II}.

\subsection{Complex bion amplitude versus $[\I\bar \I]$ amplitude}
\label{sec:cb}

The SG system also has exact complex bion solutions \cite{Behtash:2015zha,
Behtash:2015loa}. The complex bion configuration $[\mathcal C \mathcal B]$ 
is an exact solution of the Euclidean equations of motions  \eqref{PLW-3} 
derived from the quantum potential \eqref{Gen-NP-3}. In this section we 
explain its relation to the  $[\I \bar\I]$ amplitude \eqref{eq:IIbar} 
computed in Section \ref{sec:iibar-thimble}, obtained using just the 
bosonic action. Additional novel features appear in the case of $[\I\bar{\I}]$ 
pairs, beyond those  discussed in the previous section for the case of 
the $[\I{\I}]$ amplitude. 

\begin{itemize}
\item  In the $[\I\bar\I]_{\pm}$ analysis the HTA arises from the 
integration over the QZM-cycle. In the $[\cal CB]_{\pm}$ calculation
the HTA appears as the imaginary part of the action. 

\item  In  instanton calculus the Jacobian and the fluctuation determinant 
are manifestly real. In the complex bion analysis,
the Jacobian and fluctuation determinant could potentially modify the HTA 
arising from the classical action. We will see that this is not the case.  
The complex action in the Jacobian cancels exactly against a similar 
factor that appears in the GY method of calculating the determinant 
$ \left[\hat{\det}\,\mathcal{M}_{\rm complex\, bion}\right]^{-1/2}$. 

\end{itemize}

The complex bion fluctuation determinant is computed in  Appendix 
\ref{sec:det-bounce-complex-bion}. We obtain the counterpart of the 
real bion result \eqref{realbion}, in the  weak coupling limit, as 
\begin{align}
  [{\cal CB}]_{\pm}  =  \frac{1}{2\pi}   \left( {\frac{g}{32}} \right)^{+p-1}   
   \left( {\frac{p}{e}} \right)^{+p}    \sqrt {\frac {2 \pi}{p}} 
        e^{-2S_{I }  } e^{  \pm i p \pi} \
   =  [{\cal RB}]  e^{  \pm i p \pi} \, . 
\end{align}
This result shows that, up to the hidden topological phase, the one-loop 
amplitudes for the complex and the real bion cases are equal. Crucially, the 
complex bion is a point lying on the QZM-thimble of the 
instanton/anti-instanton critical point at infinity: $[\I\bar\I]_{\pm}$. 
This explains why the two types of computations are related. This thimble 
is shown in Fig.~\ref{fig:iibar-thimble}, and the complex bion configuration 
corresponds to a complex separation between the instanton and anti-instanton 
constituents given by
\begin{align}
\label{cbion}  
\tau^{*}_\pm  =\left( \ln \left( \frac {A}{gp} \right) \pm i \pi \right) \, . 
\end{align} 
Correcting for the soft mode that arises from the complex bion analysis, 
one finds precisely the $[\I\bar\I]_{\pm}$ amplitude in \eqref{eq:IIbar}, 
obtained over the QZM-thimble of the critical point at infinity.

\section{Conclusions}
\label{sec_sum}

 In this work we clarified the treatment of multi-instanton contributions
in the quantum mechanical path integral. Our main findings are:

\begin{itemize}
\item[{\bf 1)}] Generic multi-instanton contributions correspond to
critical points at infinity, and the correct way to compute them is 
to calculate the integral over the associated Lefschetz thimble. 
Typically, the main contribution comes not from the saddle point 
or its vicinity, but from a non-Gaussian integral on the associated 
complexified quasi-zero mode manifold.
 
\item[{\bf 2)}] In the theories with fermions the leading contribution
is captured by an exact solution of the equations of motion in the 
quantum modified potential. We showed how to compute the fluctuation
operators around these solutions. The fluctuation operator encodes
the exact asymptotic behavior of the thimble integral. In order
to reproduce the full result the non-Gaussian mode has to be treated 
exactly, not just in a Gaussian approximation.

\item[{\bf 3)}] We demonstrated that the fluctuation operator can
be computed for both real and complex (singular) bion solutions. The 
result has the correct form to match the thimble integration. In
particular, the cancellations between real bions and complex
singular bions that are required by supersymmetry are preserved 
when fluctuations are included.

\end{itemize}

 There are some obvious directions in which the present study 
can be extended. Within quantum mechanics, it will be interesting 
to study the thimbles that appear in correlated multi-instanton 
events beyond second order. In quantum field theory, there are 
applications to the non-BPS multi-instanton and multi-monopole 
amplitudes discussed in \cite{Unsal:2007jx,Poppitz:2012nz,Dabrowski:2013kba,
Nitta:2014vpa,Liu:2015ufa,Nitta:2015tua}.

\acknowledgments  
M.\"U. thanks Nick Manton for discussions. 
We thank the KITP at UC Santa Barbara for its hospitality 
during the program ``Resurgent Asymptotics in Physics and Mathematics'' 
where some of this work was done. Research at KITP is supported by the 
National Science Foundation under Grant No. NSF PHY-1125915. This material 
is based upon work supported by the U.S. Department of Energy, Office of 
Science, Office of High Energy Physics under Award Number DE-SC0010339 
(GD), and  by the U.S. Department of Energy, Office of Science, Office of 
Nuclear Physics under Award Number DE-FG02-03ER41260 (AB,TSc,M\"U).
 
\appendix

\section{Fluctuation Determinants}
\label{app}

In this Appendix we record the computations of the fluctuation determinants 
around the bion configurations discussed in Section \ref{sec:bions}. We 
compute the determinant with the zero mode removed using an appropriate 
variant \cite{McKane:1995vp,Kirsten:2001wz,Dunne:2005rt,Dunne:2007rt} 
of the general determinant method of Gel'fand-Yaglom
\cite{Gelfand:1959nq,Coleman:1977py,Callan:1977pt,Kirsten:2001wz,
Kleinert:2004ev,Dunne:2007rt,Marino:2015yie}. In this approach, the 
determinant with zero mode removed is expressed entirely in terms of 
the associated classical action and the asymptotic data of the zero mode as
\begin{equation}
\frac{\hat{\det}\,\mathcal{M}}{{\det}\,\mathcal{M}_0} =  
\frac{g }{2\omega A_+A_-}\, S_{\rm saddle}\, , 
\label{eq:GY}
\end{equation}
where the fluctuation operator about a SG saddle solution is given as
\begin{eqnarray}
{\mathcal M}_{\rm saddle}&=& -\frac{d^2}{dt^2} 
  + V^{\prime\prime}(z_{\rm saddle}(t))  \nonumber\\
&=& -\frac{d^2}{dt^2} +  \cos\left(z_{\rm saddle}(t)\right)
 +\frac{p\, g}{8}\cos\left(\frac{1}{2} z_{\rm saddle}(t)\right)\, . 
\label{eq:sg-saddle-fluctuation}
\end{eqnarray}
and the zero mode is the derivative of the classical saddle solution, with 
asymptotics
\begin{equation}
\dot z(t) \approx A_{\pm} e^{\mp \omega t }  \qquad {\rm for} \; 
t \rightarrow \pm \infty \, .
\label{eq:asymptotic_sol}
\end{equation}
The determinant is normalized by the determinant of the corresponding free 
operator with frequency $\omega$, $\mathcal{M}_0=-\frac{d^2}{dt^2}+\omega^2$.
For fluctuation operators about instantons, this is all very well known. For 
the SG system, the corresponding instanton zero mode is
\begin{eqnarray}
\dot{x}_I&=&2\, {\rm sech}(t) \sim 4 \, e^{\mp  t} 
  \quad, \quad t\to\pm \infty\, , 
\end{eqnarray}
from which the result quoted in \eqref{eq:sg-idet} follows. We now apply the general 
formula \eqref{eq:GY} to the results of \cite{Behtash:2015loa,Behtash:2015zha}, 
to compute the fluctuation determinants for the various bion solutions of the 
SG system.

\subsection{Fluctuation Determinant for Real Bion Saddle}
\label{sec:det-real-bion}

\begin{figure}[t]
\centerline{\includegraphics[scale=0.5]{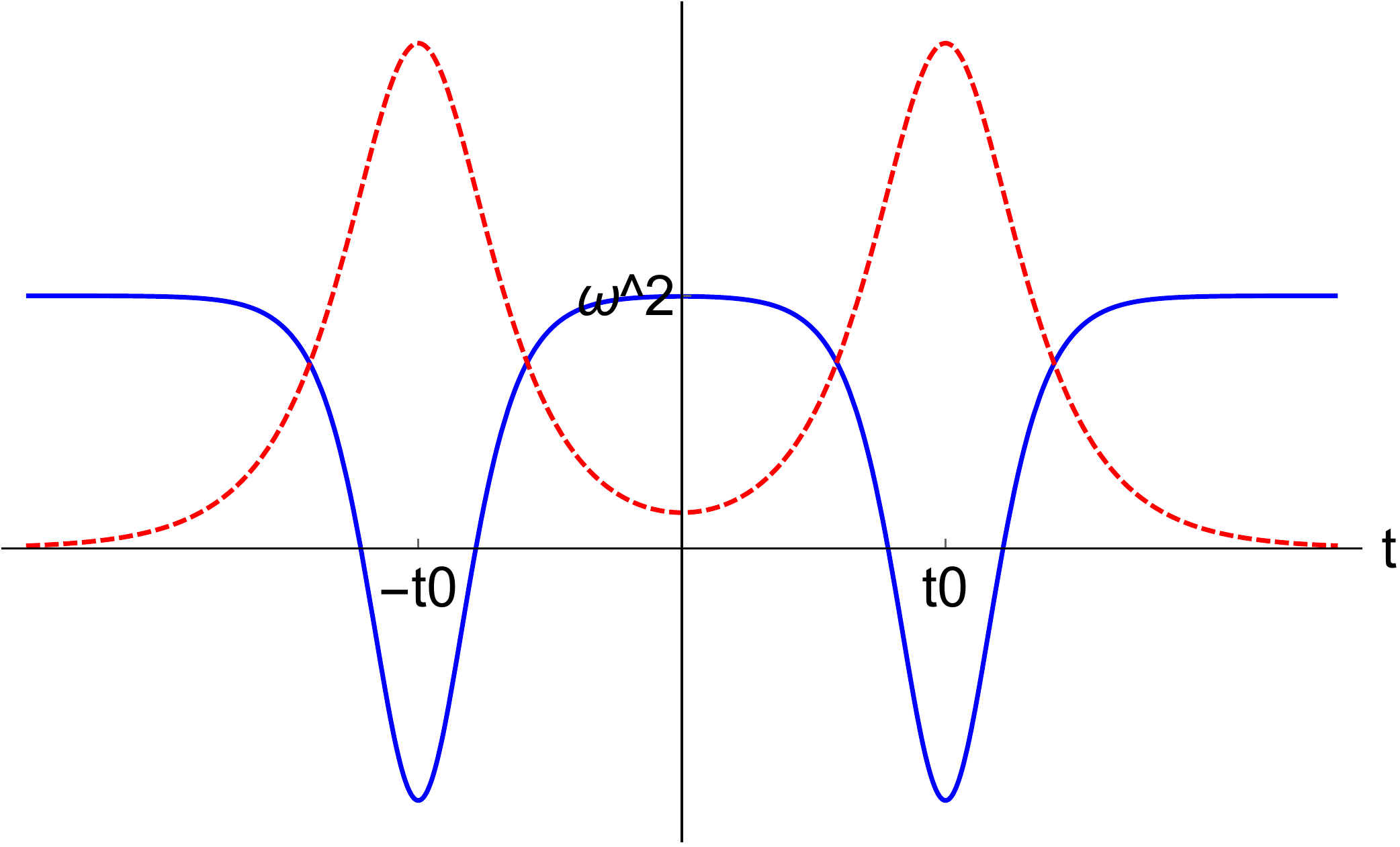}}
\caption{
Plot of the fluctuation potential in \eqref{eq:sg-real-bion-fluctuation} 
for the SG real bion solution (solid blue curve), together with the zero 
mode (dashed red curve) in \eqref{eq:sg-real-bion-zm}. Note the double 
well structure of the fluctuation potential, and the symmetric nature of 
the zero mode solution, characteristic of a real bion solution, leading 
to a positive fluctuation determinant.}
\label{fig:sg-real-bion-fluc}
\end{figure}

After integrating out $p$ flavors of fermions, the SG bosonic quantum potential 
is
\begin{eqnarray}
V(x)=\frac{1}{2}\left(W^\prime\right)^2+\frac{1}{2} p\, g\, W^{\prime\prime} 
= 2\, \sin^2\left(\frac{x}{2}\right)
  -\frac{1}{2} p\,g\, \cos\left(\frac{x}{2}\right) \, . 
\label{eq:sg-potential}
\end{eqnarray}
The complexified classical equations of motion are
\begin{eqnarray}
\ddot{z}= \sin\left(z\right)
   +\frac{1}{4} g\, p\, \sin\left(\frac{z}{2}\right) \, . 
\label{eq:sg-eom}
\end{eqnarray}
The  equations of motion \eqref{eq:sg-eom} have a real bion solution 
\cite{Behtash:2015zha,Behtash:2015loa}, which has the form of a correlated 
instanton-instanton pair,
\begin{eqnarray}
z_{\rm real\, bion}(t)
 =   4\left( {\rm arctan}\left(\exp\left[\omega(t-t_c-t_0)\right]\right) 
          +{\rm arctan}\left(\exp\left[ \omega(t-t_c+t_0)\right]\right)
  \right)\, , 
\label{eq:sg-real-bion}
\end{eqnarray}
where the frequency $\omega=\omega_{\rm real\, bion}$ is
\begin{eqnarray}
\omega_{\rm real\, bion}=   \sqrt{1 +\tfrac{p\,g}{8}}\, . 
\label{eq:sg-omega} 
\end{eqnarray}
The translational collective coordinate of the bion is $t_c$, and 
$t_0$ is the characteristic separation scale 
\begin{eqnarray}
t_0^{\rm real\, bion} \equiv \frac{1}{\omega_{\rm real\, bion} }
  \ln \left[\sqrt{\frac{8}{p\, g}}\left(1+\omega_{\rm real\, bion} \right)\right]
    \approx \frac{1}{2} \ln \left(\frac{32}{p\, g}\right)+\dots \, . 
\label{eq:sg-real-bion-t0}
\end{eqnarray} 
The action of the SG real bion solution is
\begin{eqnarray}
 S_{\rm real\, bion} 
&=& \tfrac{16}{g} \sqrt{1+\tfrac{g\, p }{8} } 
   + 2\, p\,  {\rm arctanh}\left(\frac{1}{\sqrt{1+\frac{p\, g}{8}}}\right) 
\nonumber \\
 &=& \frac{16}{g} + p \ln\left(\frac{32\, e}{p\,g}\right )+\dots \, . 
\label{eq:sg-action-real-bion}
\end{eqnarray}
The fluctuation operator about the SG real bion solution is 
\begin{eqnarray}
{\mathcal M}_{\rm real\, bion}
&=& -\frac{d^2}{dt^2} +  \cos\left(z_{\rm real\, bion}(t)\right)
 +\frac{pg}{8}\cos\left(\frac{1}{2} z_{\rm real\, bion}(t)\right)\, . 
\label{eq:sg-real-bion-fluctuation}
\end{eqnarray}
The form of this fluctuation potential is plotted in 
Fig.~\ref{fig:sg-real-bion-fluc}.

The fluctuation operator in \eqref{eq:sg-real-bion-fluctuation} has an 
exact zero mode, associated with translation symmetry
\begin{eqnarray}
\phi^{\rm real\, bion}_\text{zero mode}(t)&=& \dot{z}_{\rm real\, bion}(t)
  \nonumber\\
  &=& 2\, \omega \left({\rm sech}\left(\omega(t-t_c-t_0)\right) 
  +{\rm sech}\left(\omega(t-t_c+t_0)\right) \right)\, ,
\label{eq:sg-real-bion-zm}
\end{eqnarray}
where $\omega\equiv \omega_{\rm real\, bion}$, and $t_0\equiv t_0^{\rm real\, bion}$. 
This  zero mode \eqref{eq:sg-real-bion-zm} is also plotted in 
Fig.~\ref{fig:sg-real-bion-fluc}. Note the double-well structure of the 
fluctuation potential, distinct from the familiar single-well structure 
of the fluctuation potential \eqref{eq:sg-ifluc} for the instanton solution. 
Also note the symmetric form of the real bion zero mode, as expected for 
the lowest mode of a non-negative fluctuation operator (contrast with the 
anti-symmetric zero mode for the bounce, in Fig.~\ref{fig:sg-bounce-fluc}).

To compute the real bion fluctuation determinant we need the asymptotic values of 
the zero mode as normalized in \eqref{eq:sg-real-bion-zm}. For the SG 
real bion solution
\begin{eqnarray}
\phi^{\rm real\, bion}_\text{zero mode}(t)
 \sim 4  \sqrt{\frac{32}{p\, g}} \,\omega^2_{\rm real\, bion}
   e^{\mp \omega_{\rm real\, bion} \, t}  
\quad, \quad t\to\pm \infty\, .
\label{eq:sg-real-bion-zm-asymptotics}
\end{eqnarray}
Thus, from the general formula \eqref{eq:GY}, the determinant of the 
fluctuation operator with zero mode removed, relative to that of the 
corresponding free operator, is given by
\begin{eqnarray}
 \frac{\hat{\det}\,\mathcal{M}_{\rm real\, bion}}
  {{\det}\,\mathcal{M}_0} 
&  \approx  & 
\frac{p\, g}{64}\, . 
\label{eq:sg-real-bion-det}
\end{eqnarray}

\subsection{Fluctuation Determinants for Real Bounce and Complex Bion Saddles}
\label{sec:det-bounce-complex-bion}

We first recall the real bounce solution, because as discussed in detail 
in \cite{Behtash:2015loa,Behtash:2015zha}, the complex bion solutions are 
obtained from the real bounce by the analytic continuation $p\to -p$. We thus compute 
the fluctuation determinants  from the analytic continuation of the 
corresponding real bounce fluctuation determinant, as all steps of the 
computation have a well-defined analytic continuation. We will see that this procedure leads to results that agree perfectly with the physical picture developed in this paper.

\begin{figure}[t]
\centerline{\includegraphics[scale=0.5]{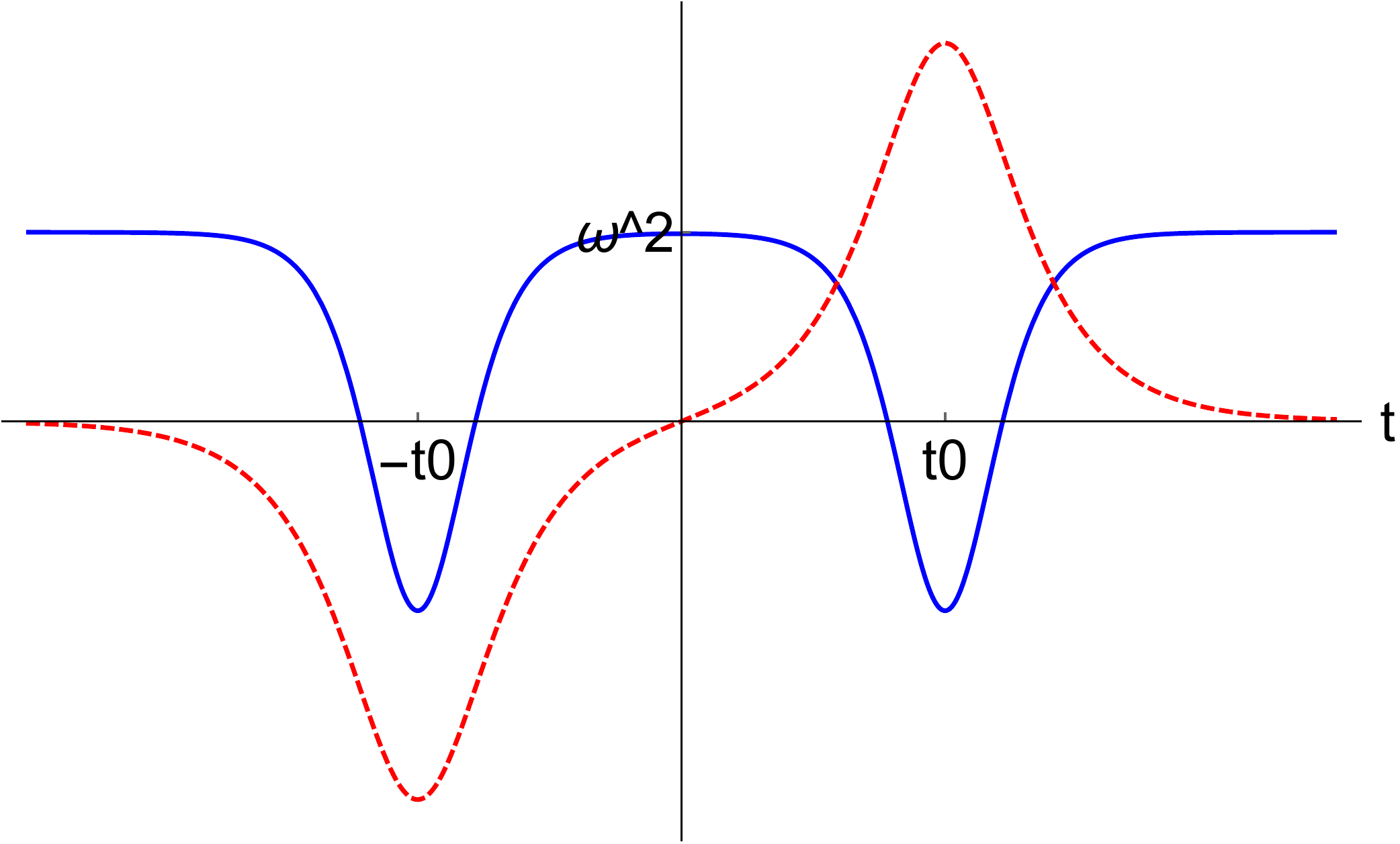}}
\caption{Plot of the fluctuation potential in \eqref{eq:sg-bounce-fluctuation} 
for the SG bounce solution (solid blue curve), together with the zero mode 
(dashed red curve) in \eqref{eq:sg-bounce-zm}. Note the double-well structure 
of the fluctuation potential, and the anti-symmetric nature of the zero mode 
solution, characteristic of a bounce solution, leading to a negative 
fluctuation determinant.}
\label{fig:sg-bounce-fluc}
\end{figure}

The  equations of motion \eqref{eq:sg-eom} have a real bounce solution, with 
the form of a correlated instanton/anti-instanton pair
\begin{eqnarray}
z_{\rm bounce}(t)
 = 2\pi  +4\Big[ {\rm arctan}\left(\exp\left[ \omega(t-t_c-t_0)\right]\right) 
  - {\rm arctan}\left(\exp\left[ \omega(t-t_c+t_0)\right]\right)\Big]\, , 
  \nonumber\\
\label{eq:sg-bounce}
\end{eqnarray}
The frequency $\omega$ in \eqref{eq:sg-bounce} is
\begin{eqnarray}
\omega_{\rm bounce}=   \sqrt{1 -\tfrac{p\,g}{8}}
\label{eq:sg-b-omega} 
\end{eqnarray}
and the characteristic separation scale $t_0$ in \eqref{eq:sg-bounce} is
\begin{eqnarray}
t_0^{\rm bounce} \equiv \frac{1}{\omega_{\rm bounce} }
  \ln \left[\sqrt{\frac{8}{p\, g}}\left(1+\omega_{\rm bounce} \right)\right]
    \approx \frac{1}{2} \ln \left(\frac{32}{p\, g}\right)+\dots \, . 
    \label{eq:sg-t0}
\end{eqnarray}
The action of the SG bounce solution is
\begin{eqnarray}
 S_{\rm bounce} 
 &=&  \tfrac{16}{g} \sqrt{1-\tfrac{g \,p }{8} } 
     - 2 p\,  {\rm arctanh} \left[\sqrt{1-\tfrac{g\, p}{8}}\right ] 
\nonumber \\
  &=& \frac{16}{g} - p\ln\left(\frac{32\, e}{p\, g}\right )+\dots \, . 
  \label{eq:sg-bounce-action}
\end{eqnarray}
The fluctuation operator about this SG bounce solution is
\begin{eqnarray}
{\mathcal M}_{\rm bounce}
&=& -\frac{d^2}{dt^2} +  \cos\left(z_{\rm bounce}(t)\right)
  +\frac{pg}{8}\cos\left(\frac{1}{2} z_{\rm bounce}(t)\right)\, . 
\label{eq:sg-bounce-fluctuation}
\end{eqnarray}
The form of this fluctuation potential is plotted in 
Fig.~\ref{fig:sg-bounce-fluc}.

This fluctuation operator in \eqref{eq:sg-bounce-fluctuation} has an 
exact zero mode, associated with translation symmetry
\begin{eqnarray}
\phi^{\rm bounce}_\text{zero mode}(t)&=& \dot{z}_{\rm bounce}(t)
\nonumber\\
  &=& 2\, \omega \left({\rm sech}\left(\omega(t-t_c-t_0)\right) 
  - {\rm sech}\left(\omega(t-t_c+t_0)\right) \right)\, , 
  \label{eq:sg-bounce-zm}
\end{eqnarray}
where $\omega\equiv \omega_{\rm bounce}$, and $t_0\equiv t_0^{\rm bounce}$.
This SG bounce  zero mode \eqref{eq:sg-bounce-zm} is also plotted in 
Fig.~\ref{fig:sg-bounce-fluc}.  Note the double-well structure, distinct 
from the familiar single-well structure of the fluctuations about the 
instanton solution. Also note the anti-symmetric form of the bounce 
zero mode, indicating the existence of a negative mode, as expected 
for a bounce solution (contrast with the symmetric zero mode for the 
real bion in Fig.~\ref{fig:sg-real-bion-fluc}).

To compute 
the fluctuation determinant we need the asymptotic values of the zero mode,
as normalized in \eqref{eq:sg-bounce-zm}. For this SG bounce solution
\begin{eqnarray}
\phi^{\rm bounce}_\text{zero mode}(t)
  \sim \pm 8  \sqrt{\frac{8}{p\, g}} \, \omega^2_{\rm bounce} 
    e^{\mp \, \omega_{\rm bounce} \, t}
  \quad, \quad t\to\pm \infty\, . 
\label{eq:sg-bounce-zm-asymptotics}
\end{eqnarray}
Thus, from the general formula \eqref{eq:GY}, the determinant of the bounce 
fluctuation operator with zero mode removed, relative to that of the 
corresponding free operator, is given by
\begin{eqnarray}
 \frac{\hat{\det}\,\mathcal{M}_{\rm bounce}}
                             {{\det}\,\mathcal{M}_0} 
&   \approx & -\frac{p\, g}{64}\, . 
\label{eq:sg-bounce-det}
\end{eqnarray}
The supersymmetric SG system also has a complex bion solution, from the true 
vacuum critical point to (a complex conjugate pair of) complex turning points. 
As discussed in detail in \cite{Behtash:2015loa,Behtash:2015zha}, this solution 
is obtained by implementing the analytic continuation $p\to e^{\pm i \pi}\, p$ 
in the (real) bounce solution. Thus, by an analogous computation, the 
corresponding  fluctuation determinant is
\begin{eqnarray}
 \frac{\hat{\det}\,\mathcal{M}_{\rm complex\, bion}}
                                   {{\det}\,\mathcal{M}_0} 
&  \approx  & 
\frac{p\, g}{64}\, . 
\label{eq:sg-complex-bion-det}
\end{eqnarray}

\bibliography{bibliography_NonGaussian}

\providecommand{\href}[2]{#2}\begingroup\raggedright\begin{thebibliography}{10}

\bibitem{Coleman:1978ae}
S.~R. Coleman, \emph{Aspects of Symmetry}.
\newblock Cambridge University Press, 1979.

\bibitem{ZinnJustin:2002ru}
J.~Zinn-Justin, \emph{{Quantum field theory and critical phenomena}},
  {\emph{Int.Ser.Monogr.Phys.} {\bf 113} (2002) 1--1054}.

\bibitem{Bogomolny:1980ur}
E.~Bogomolny, \emph{{Calculation of instanton-anti-instanton contributions in
  quantum mechanics}},
  \href{http://dx.doi.org/10.1016/0370-2693(80)91014-X}{\emph{Phys.Lett.} {\bf
  B91} (1980) 431--435}.

\bibitem{ZinnJustin:1981dx}
J.~Zinn-Justin, \emph{{Multi - Instanton Contributions in Quantum Mechanics}},
  \href{http://dx.doi.org/10.1016/0550-3213(81)90197-8}{\emph{Nucl.Phys.} {\bf
  B192} (1981) 125--140}.

\bibitem{Cherman:2014ofa}
A.~Cherman, D.~Dorigoni and M.~Unsal, \emph{{Decoding perturbation theory using
  resurgence: Stokes phenomena, new saddle points and Lefschetz thimbles}},
  \href{http://dx.doi.org/10.1007/JHEP10(2015)056}{\emph{JHEP} {\bf 10} (2015)
  056}, [\href{http://arxiv.org/abs/1403.1277}{{\tt 1403.1277}}].

\bibitem{Dunne:2012ae}
G.~V. Dunne and M.~{\"U}nsal, \emph{{Resurgence and Trans-series in Quantum
  Field Theory: The CP(N-1) Model}},
  \href{http://dx.doi.org/10.1007/JHEP11(2012)170}{\emph{JHEP} {\bf 1211}
  (2012) 170}, [\href{http://arxiv.org/abs/1210.2423}{{\tt 1210.2423}}].

\bibitem{Cherman:2013yfa}
A.~Cherman, D.~Dorigoni, G.~V. Dunne and M.~{\"U}nsal, \emph{{Resurgence in
  Quantum Field Theory: Nonperturbative Effects in the Principal Chiral
  Model}},
  \href{http://dx.doi.org/10.1103/PhysRevLett.112.021601}{\emph{Phys.Rev.Lett.}
  {\bf 112} (2014) 021601}, [\href{http://arxiv.org/abs/1308.0127}{{\tt
  1308.0127}}].

\bibitem{Misumi:2015dua}
T.~Misumi, M.~Nitta and N.~Sakai, \emph{{Resurgence in sine-Gordon quantum
  mechanics: Exact agreement between multi-instantons and uniform WKB}},
  \href{http://arxiv.org/abs/1507.00408}{{\tt 1507.00408}}.

\bibitem{Basar:2013eka}
G.~Basar, G.~V. Dunne and M.~{\"U}nsal, \emph{{Resurgence theory,
  ghost-instantons, and analytic continuation of path integrals}},
  \href{http://dx.doi.org/10.1007/JHEP10(2013)041}{\emph{JHEP} {\bf 10} (2013)
  041}, [\href{http://arxiv.org/abs/1308.1108}{{\tt 1308.1108}}].

\bibitem{Argyres:2012ka}
P.~C. Argyres and M.~{\"U}nsal, \emph{{The semi-classical expansion and
  resurgence in gauge theories: new perturbative, instanton, bion, and
  renormalon effects}},
  \href{http://dx.doi.org/10.1007/JHEP08(2012)063}{\emph{JHEP} {\bf 1208}
  (2012) 063}, [\href{http://arxiv.org/abs/1206.1890}{{\tt 1206.1890}}].

\bibitem{Anber:2014sda}
M.~M. Anber and T.~Sulejmanpasic, \emph{{The absence of IR renormalons in gauge
  theories on $\mathbb R^3\times \mathbb S^1$ and what it means for
  resurgence}},  \href{http://arxiv.org/abs/1410.0121}{{\tt 1410.0121}}.

\bibitem{Behtash:2015loa}
A.~Behtash, G.~V. Dunne, T.~Sch\"afer, T.~Sulejmanpasic and M.~Unsal,
  \emph{{Toward Picard-Lefschetz Theory of Path Integrals, Complex Saddles and
  Resurgence}}, \href{http://dx.doi.org/10.4310/AMSA.2017.v2.n1.a3}{\emph{Ann.
  Math. Sci. and Appl.} {\bf 2} (2015) 95--212},
  [\href{http://arxiv.org/abs/1510.03435}{{\tt 1510.03435}}].

\bibitem{Dunne:2016nmc}
G.~V. Dunne and M.~{\"U}nsal, \emph{{New Nonperturbative Methods in Quantum
  Field Theory: From Large-N Orbifold Equivalence to Bions and Resurgence}},
  \href{http://dx.doi.org/10.1146/annurev-nucl-102115-044755}{\emph{Ann. Rev.
  Nucl. Part. Sci.} {\bf 66} (2016) 245--272},
  [\href{http://arxiv.org/abs/1601.03414}{{\tt 1601.03414}}].

\bibitem{Sulejmanpasic:2016fwr}
T.~Sulejmanpasic and M.~{\"U}nsal, \emph{{Aspects of Perturbation theory in
  Quantum Mechanics: The BenderWu Mathematica package}},
  \href{http://arxiv.org/abs/1608.08256}{{\tt 1608.08256}}.

\bibitem{Kontsevich-3}
M.~Kontsevich, \emph{On non-perturbative quantization, fukaya categories and
  resurgence}, {\emph{Talk at Simons Center} (2015) }.

\bibitem{Kontsevich-1}
M.~Kontsevich, \emph{Resurgence from the path integral perspective},
  {\emph{Talk at Perimeter Institute} (2012) }.

\bibitem{Witten:2010zr}
E.~Witten, \emph{{A New Look At The Path Integral Of Quantum Mechanics}},
  \href{http://arxiv.org/abs/1009.6032}{{\tt 1009.6032}}.

\bibitem{Witten:2010cx}
E.~Witten, \emph{{Analytic Continuation Of Chern-Simons Theory}}, {\emph{AMS/IP
  Stud. Adv. Math.} {\bf 50} (2011) 347--446},
  [\href{http://arxiv.org/abs/1001.2933}{{\tt 1001.2933}}].

\bibitem{Harlow:2011ny}
D.~Harlow, J.~Maltz and E.~Witten, \emph{{Analytic Continuation of Liouville
  Theory}}, \href{http://dx.doi.org/10.1007/JHEP12(2011)071}{\emph{JHEP} {\bf
  1112} (2011) 071}, [\href{http://arxiv.org/abs/1108.4417}{{\tt 1108.4417}}].

\bibitem{Behtash:2015kva}
A.~Behtash, E.~Poppitz, T.~Sulejmanpasic and M.~{\"U}nsal, \emph{{The curious
  incident of multi-instantons and the necessity of Lefschetz thimbles}},
  \href{http://arxiv.org/abs/1507.04063}{{\tt 1507.04063}}.

\bibitem{Behtash:2017rqj}
A.~Behtash, \emph{{More on Homological Supersymmetric Quantum Mechanics}},
  \href{http://dx.doi.org/10.1103/PhysRevD.97.065002}{\emph{Phys.Rev.} {\bf
  D97} (2018) 065002}, [\href{http://arxiv.org/abs/1703.00511}{{\tt
  1703.00511}}].

\bibitem{Behtash:2015kna}
A.~Behtash, T.~Sulejmanpasic, T.~Sch\"afer and M.~\"Unsal, \emph{{Hidden
  Topological Angles in Path Integrals}},
  \href{http://dx.doi.org/10.1103/PhysRevLett.115.041601}{\emph{Phys. Rev.
  Lett.} {\bf 115} (2015) 041601}, [\href{http://arxiv.org/abs/1502.06624}{{\tt
  1502.06624}}].

\bibitem{Behtash:2015zha}
A.~Behtash, G.~V. Dunne, T.~Sch{\"a}fer, T.~Sulejmanpasic and M.~{\"U}nsal,
  \emph{{Complexified path integrals, exact saddles and supersymmetry}},
  \href{http://dx.doi.org/10.1103/PhysRevLett.116.011601}{\emph{Phys. Rev.
  Lett.} {\bf 116} (2016) 011601}, [\href{http://arxiv.org/abs/1510.00978}{{\tt
  1510.00978}}].

\bibitem{Kozcaz:2016wvy}
C.~Koz{\c c}az, T.~Sulejmanpasic, Y.~Tanizaki and M.~{\"U}nsal, \emph{{Cheshire
  Cat resurgence, Self-resurgence and Quasi-Exact Solvable Systems}},
  \href{http://arxiv.org/abs/1609.06198}{{\tt 1609.06198}}.

\bibitem{Fujimori:2016ljw}
T.~Fujimori, S.~Kamata, T.~Misumi, M.~Nitta and N.~Sakai,
  \emph{{Nonperturbative contributions from complexified solutions in
  $\mathbb{C}P^{N-1}$models}},
  \href{http://dx.doi.org/10.1103/PhysRevD.94.105002}{\emph{Phys. Rev.} {\bf
  D94} (2016) 105002}, [\href{http://arxiv.org/abs/1607.04205}{{\tt
  1607.04205}}].

\bibitem{Dorigoni:2017smz}
D.~Dorigoni and P.~Glass, \emph{{The grin of Cheshire cat resurgence from
  supersymmetric localization}},  \href{http://arxiv.org/abs/1711.04802}{{\tt
  1711.04802}}.

\bibitem{Nekrasov:2018pqq}
N.~Nekrasov, \emph{{Tying up instantons with anti-instantons}},
  \href{http://arxiv.org/abs/1802.04202}{{\tt 1802.04202}}.

\bibitem{fedoryuk}
M.~V. Fedoryuk, \emph{The saddle-point method}, {\emph{Izdat. ``Nauka,''
  Moscow, MR 58:22580} (1977) }.

\bibitem{Schafer:1996wv}
T.~Sch{\"a}fer and E.~V. Shuryak, \emph{{Instantons in QCD}},
  \href{http://dx.doi.org/10.1103/RevModPhys.70.323}{\emph{Rev.Mod.Phys.} {\bf
  70} (1998) 323--426}, [\href{http://arxiv.org/abs/hep-ph/9610451}{{\tt
  hep-ph/9610451}}].

\bibitem{Pham}
F.~Pham, \emph{Vanishing homologies and the n variable saddlepoint method},
  {\emph{Proc. Symp. Pure Math} {\bf 2} (1983) 319--333.}

\bibitem{arnold}
V.~I. Arnold, S.~M. Gusein-Zade and A.~N. Varchenko, \emph{Singularities of
  Differentiable Maps, Volume 1}.
\newblock Birkh{\"a}user Basel, 2012.

\bibitem{Witten:1982df}
E.~Witten, \emph{{Constraints on Supersymmetry Breaking}},
  \href{http://dx.doi.org/10.1016/0550-3213(82)90071-2}{\emph{Nucl.Phys.} {\bf
  B202} (1982) 253}.

\bibitem{Shifman:1995mm}
M.~A. Shifman, \emph{{Beginning supersymmetry (supersymmetry in quantum
  mechanics). In Shifman, M.A.: {\it ITEP lectures on particle physics and
  field theory}, Vol. 1, pp 301-344.}}
\newblock World Scientific (Singapore), 1995.

\bibitem{Balitsky:1985in}
I.~Balitsky and A.~Yung, \emph{{Instanton Molecular Vacuum in $N=1$
  Supersymmetric Quantum Mechanics}},
  \href{http://dx.doi.org/10.1016/0550-3213(86)90295-6}{\emph{Nucl.Phys.} {\bf
  B274} (1986) 475}.

\bibitem{Buividovich:2015oju}
P.~V. Buividovich, G.~V. Dunne and S.~N. Valgushev, \emph{{Complex Path
  Integrals and Saddles in Two-Dimensional Gauge Theory}},
  \href{http://dx.doi.org/10.1103/PhysRevLett.116.132001}{\emph{Phys. Rev.
  Lett.} {\bf 116} (2016) 132001}, [\href{http://arxiv.org/abs/1512.09021}{{\tt
  1512.09021}}].

\bibitem{Serone:2016qog}
M.~Serone, G.~Spada and G.~Villadoro, \emph{{Instantons from Perturbation
  Theory}}, \href{http://dx.doi.org/10.1103/PhysRevD.96.021701}{\emph{Phys.
  Rev.} {\bf D96} (2017) 021701}, [\href{http://arxiv.org/abs/1612.04376}{{\tt
  1612.04376}}].

\bibitem{Serone:2017nmd}
M.~Serone, G.~Spada and G.~Villadoro, \emph{{The Power of Perturbation
  Theory}}, \href{http://dx.doi.org/10.1007/JHEP05(2017)056}{\emph{JHEP} {\bf
  05} (2017) 056}, [\href{http://arxiv.org/abs/1702.04148}{{\tt 1702.04148}}].

\bibitem{DST}
{Gerald Dunne, Tin Sulejmanpasic and Mithat {\"U}nsal, {\it Work in Progress}}.

\bibitem{Brezin:1976wa}
E.~Brezin, J.-C. Le~Guillou and J.~Zinn-Justin, \emph{{Perturbation Theory at
  Large Order. 2. Role of the Vacuum Instability}},
  \href{http://dx.doi.org/10.1103/PhysRevD.15.1558}{\emph{Phys.Rev.} {\bf D15}
  (1977) 1558--1564}.

\bibitem{Balian:1978et}
R.~Balian, G.~Parisi and A.~Voros, \emph{{Quartic Oscillator}},  in
  \emph{{Feynman Path Integrals. Proceedings of the International Colloquium
  held in Marseille, May 1978}}, pp.~337--360, 1978.

\bibitem{Richard:1981gn}
J.~L. Richard and A.~Rouet, \emph{{Complex Saddle Points Versus Dilute Gas
  Approximation in the Double Well Anharmonic Oscillator}},
  \href{http://dx.doi.org/10.1016/0550-3213(81)90363-1}{\emph{Nucl. Phys.} {\bf
  B185} (1981) 47--60}.

\bibitem{Lapedes:1981tz}
A.~Lapedes and E.~Mottola, \emph{{Complex Path Integrals and Finite
  Temperature}},
  \href{http://dx.doi.org/10.1016/0550-3213(82)90477-1}{\emph{Nucl. Phys.} {\bf
  B203} (1982) 58}.

\bibitem{Millard:1984qt}
P.~A. Millard, \emph{{Complex Classical Paths and the One-dimensional
  {Sine-Gordon} System}},
  \href{http://dx.doi.org/10.1016/0550-3213(85)90636-4}{\emph{Nucl. Phys.} {\bf
  B259} (1985) 266}.

\bibitem{Gelfand:1959nq}
I.~M. Gelfand and A.~M. Yaglom, \emph{{Integration in functional spaces and it
  applications in quantum physics}},
  \href{http://dx.doi.org/10.1063/1.1703636}{\emph{J. Math. Phys.} {\bf 1}
  (1960) 48}.

\bibitem{Coleman:1977py}
S.~R. Coleman, \emph{{The Fate of the False Vacuum. 1. Semiclassical Theory}},
  \href{http://dx.doi.org/10.1103/PhysRevD.15.2929,
  10.1103/PhysRevD.16.1248}{\emph{Phys. Rev.} {\bf D15} (1977) 2929--2936}.

\bibitem{Callan:1977pt}
C.~G. Callan, Jr. and S.~R. Coleman, \emph{{The Fate of the False Vacuum. 2.
  First Quantum Corrections}},
  \href{http://dx.doi.org/10.1103/PhysRevD.16.1762}{\emph{Phys. Rev.} {\bf D16}
  (1977) 1762--1768}.

\bibitem{Kirsten:2001wz}
K.~Kirsten, \emph{{Spectral functions in mathematics and physics}}.
\newblock Chapman and Hall/CRC, 2001.

\bibitem{Kleinert:2004ev}
H.~Kleinert, \emph{Path Integrals in Quantum Mechanics, Statistics, Polymer
  Physics, and Financial Markets}.
\newblock World Scientific, 2006.

\bibitem{Dunne:2007rt}
G.~V. Dunne, \emph{{Functional determinants in quantum field theory}},
  \href{http://dx.doi.org/10.1088/1751-8113/41/30/304006}{\emph{J. Phys.} {\bf
  A41} (2008) 304006}, [\href{http://arxiv.org/abs/0711.1178}{{\tt
  0711.1178}}].

\bibitem{Marino:2015yie}
M.~Marino, \emph{{Instantons and Large N}}.
\newblock Cambridge University Press, 2015.

\bibitem{McKane:1995vp}
A.~J. McKane and M.~B. Tarlie, \emph{{Regularization of functional determinants
  using boundary perturbations}},
  \href{http://dx.doi.org/10.1088/0305-4470/28/23/032}{\emph{J. Phys.} {\bf
  A28} (1995) 6931--6942}, [\href{http://arxiv.org/abs/cond-mat/9509126}{{\tt
  cond-mat/9509126}}].

\bibitem{Dunne:2005rt}
G.~V. Dunne and H.~Min, \emph{{Beyond the thin-wall approximation: Precise
  numerical computation of prefactors in false vacuum decay}},
  \href{http://dx.doi.org/10.1103/PhysRevD.72.125004}{\emph{Phys. Rev.} {\bf
  D72} (2005) 125004}, [\href{http://arxiv.org/abs/hep-th/0511156}{{\tt
  hep-th/0511156}}].

\bibitem{friedli_velenik_2017}
S.~Friedli and Y.~Velenik, \emph{Statistical Mechanics of Lattice Systems: A
  Concrete Mathematical Introduction}.
\newblock Cambridge University Press, 2017,
  \href{http://dx.doi.org/10.1017/9781316882603}{10.1017/9781316882603}.

\bibitem{Dunne:2016jsr}
G.~V. Dunne and M.~Unsal, \emph{{Deconstructing zero: resurgence, supersymmetry
  and complex saddles}},
  \href{http://dx.doi.org/10.1007/JHEP12(2016)002}{\emph{JHEP} {\bf 12} (2016)
  002}, [\href{http://arxiv.org/abs/1609.05770}{{\tt 1609.05770}}].

\bibitem{Unsal:2007jx}
M.~{\"U}nsal, \emph{{Magnetic bion condensation: A New mechanism of confinement
  and mass gap in four dimensions}},
  \href{http://dx.doi.org/10.1103/PhysRevD.80.065001}{\emph{Phys.Rev.} {\bf
  D80} (2009) 065001}, [\href{http://arxiv.org/abs/0709.3269}{{\tt
  0709.3269}}].

\bibitem{Poppitz:2012nz}
E.~Poppitz, T.~Sch{\"a}fer and M.~{\"U}nsal, \emph{{Universal mechanism of
  (semi-classical) deconfinement and theta-dependence for all simple groups}},
  \href{http://dx.doi.org/10.1007/JHEP03(2013)087}{\emph{JHEP} {\bf 1303}
  (2013) 087}, [\href{http://arxiv.org/abs/1212.1238}{{\tt 1212.1238}}].

\bibitem{Dabrowski:2013kba}
R.~Dabrowski and G.~V. Dunne, \emph{{Fractionalized Non-Self-Dual Solutions in
  the CP(N-1) Model}},
  \href{http://dx.doi.org/10.1103/PhysRevD.88.025020}{\emph{Phys. Rev.} {\bf
  D88} (2013) 025020}, [\href{http://arxiv.org/abs/1306.0921}{{\tt
  1306.0921}}].

\bibitem{Nitta:2014vpa}
M.~Nitta, \emph{{Fractional instantons and bions in the O $(N)$ model with
  twisted boundary conditions}},
  \href{http://dx.doi.org/10.1007/JHEP03(2015)108}{\emph{JHEP} {\bf 03} (2015)
  108}, [\href{http://arxiv.org/abs/1412.7681}{{\tt 1412.7681}}].

\bibitem{Liu:2015ufa}
Y.~Liu, E.~Shuryak and I.~Zahed, \emph{{Confining dyon-antidyon Coulomb liquid
  model. I.}}, \href{http://dx.doi.org/10.1103/PhysRevD.92.085006}{\emph{Phys.
  Rev.} {\bf D92} (2015) 085006}, [\href{http://arxiv.org/abs/1503.03058}{{\tt
  1503.03058}}].

\bibitem{Nitta:2015tua}
M.~Nitta, \emph{{Fractional instantons and bions in the principal chiral model
  on $ {\mathrm{\mathbb{R}}}^2\times {S}^1 $ with twisted boundary
  conditions}}, \href{http://dx.doi.org/10.1007/JHEP08(2015)063}{\emph{JHEP}
  {\bf 08} (2015) 063}, [\href{http://arxiv.org/abs/1503.06336}{{\tt
  1503.06336}}].

\end{thebibliography}\endgroup
\bibliographystyle{JHEP}
\end{document}